\documentclass[12pt, letterpaper]{JHEP3}

\usepackage{amssymb, amsmath, amsopn, amsthm}
\usepackage{epsfig}

\title{Type IIA Moduli Stabilization}

\author{Oliver DeWolfe\\
Department of Physics\\
Princeton University\\
Princeton, NJ 08544, U.S.A.\\
{\tt odewolfe@princeton.edu}}

\author{Alexander Giryavets\footnote{On leave from Steklov
    Mathematical Institute, Moscow, Russia},~ Shamit Kachru\\
Department of Physics and SLAC\\
Stanford University\\
Stanford, CA 94305/94309, U.S.A.\\
{\tt giryav@stanford.edu, skachru@stanford.edu}}

\author{Washington Taylor \\
{Center for Theoretical Physics} \\ {Massachusetts Institute of Technology} \\
{Cambridge, MA 02139, U.S.A.} \\ {\tt wati\ {\rm at}\ mit.edu}}

\abstract{We demonstrate that flux compactifications of type IIA
string theory can classically stabilize all geometric moduli.  For a particular
orientifold background, we explicitly construct an infinite family of supersymmetric vacua
with all moduli stabilized at arbitrarily large volume, weak coupling, and small
negative cosmological constant. We obtain
these solutions from both ten-dimensional and four-dimensional
perspectives.  For more general backgrounds, we study the equations for
supersymmetric vacua coming from the effective superpotential and show that all
geometric moduli can be stabilized by fluxes.  We comment on the
resulting picture of statistics on the landscape of vacua.}

\preprint{hep-th/0505160, MIT-CTP-3640, PUPT-2161, SU-ITP-05/16, SLAC-PUB-11153}

\newcommand{\bC}{\mathbb{C}}
\newcommand{\bP}{\mathbb{P}}
\newcommand{\bZ}{\mathbb{Z}}
\newcommand{\bR}{\mathbb{R}}

\renewcommand{\Re}{{\rm Re}\,}
\renewcommand{\Im}{{\rm Im}\,}
\newcommand{\vol}{{\rm vol}}
\newcommand{\sgn}{{\rm sgn}\,}

\begin{document}
\section{Introduction}

It is an important problem to understand the effects which stabilize
moduli in quasi-realistic string compactifications.  Stabilized
compactifications likely provide the correct setting both for stringy models
of early universe cosmology, and
for string-based models of particle phenomenology.
In addition, the properties of the set of such vacua (perhaps endowed
with a preferred cosmological measure) may suggest new predictions,
or at least
possible interesting phenomenological signatures, of string theory.

This problem has received a great deal of attention recently.
In the framework of low-energy supersymmetry, the most concrete
constructions have appeared in the IIB theory \cite{KKLT,KKLTexample,
nonsusyKKLT}, while
constructions which break supersymmetry at a high scale have been
described in both critical and noncritical
string theories \cite{Eva}.
Proposals for constructing stabilized models in the 11D,
heterotic and type I limits have also appeared \cite{Bobby,Carlos,Heterotic,Antoniadis}.
The range of constructions seems
to be quite large, realizing the idea of a discretuum \cite{BP} and
probably requiring statistical analysis to get a reasonable picture
of the set of possibilities \cite{Douglas,statrefs,statrefsII}.

While the evidence for the existence of many stabilized vacua is quite
suggestive, it is fair to say that
it has been hard to come by extremely controlled individual examples.
The main problem is that, by definition, any
concrete example cannot have tunable couplings left over, since the
string coupling and radii have been fixed.  In the IIB context,
it has proven possible to obtain supersymmetric vacua with
weak string couplings,
and radii which grow as the logarithm of a tuning parameter \cite{KKLT};
completely explicit examples appear in \cite{KKLTexample}.
This leads to control, but only through
fine tuning by appropriate choices
in a large space of flux vacua.
For nonsupersymmetric IIB vacua, it has been argued that one can obtain
``large extra dimensions'' as well
by looking at scaling
regimes for moduli where loop and non-perturbative corrections to the
potential
conspire to make this possible \cite{nonsusyKKLT}.

In this paper, we show that it is possible to construct stabilized
vacua with arbitrarily weak coupling $g_s$ and large radius $R$ in the setting
of type IIA Calabi-Yau compactifications with flux.  We do this by
demonstrating the existence of infinite families of vacua where
$g_s$ and $R$ have power law dependence on a flux which is unconstrained
by tadpoles, and asymptote to weak coupling and large radius in the large
flux limit.  Our solutions can be seen both directly from classical 10D
supergravity and from the effective 4D framework developed in
\cite{Grimm-Louis-IIA} and extended here.
We note that it was anticipated in the papers \cite{KK,Grimm-Louis-IIA} that
generic fluxes should stabilize the geometric moduli of
IIA Calabi-Yau models,\footnote{General discussions
of IIA compactifications on spaces with
various G-structures appear in \cite{others, Generalized}.}
and in \cite{Zwirner} it was shown that untwisted moduli could be stabilized by fluxes in a particular IIA orientifold.  The main advance here is to provide an example with all moduli stabilized, to make the generic stabilization
of moduli explicit,
and to demonstrate the existence of vacua with very large radius and
weak coupling, where it is clear that all approximations are controlled.

In Section \ref{sec:model}, we introduce the simple toroidal
orientifold compactification of type IIA string theory that will be
our example.  In Section \ref{sec:10D}, we analyze this orientifold
model in the presence of fluxes using type IIA supergravity in 10
dimensions, and show that the moduli are all classically stabilized.
In Section \ref{sec:4D} we present a general analysis of IIA
compactifications from the point of view of ${\cal N}= 1$ supergravity
in 4D, extending the earlier work of Grimm and Louis
\cite{Grimm-Louis-IIA}.  We show using this formalism that the
classical stabilization of geometric moduli is generically possible in
IIA orientifold compactifications, and demonstrate the generic
existence of families of vacua admitting parametric control over the
volume and string coupling.  In Section \ref{sec:4D-applied} we apply
the general 4D analysis to the model of Section \ref{sec:model} and
relate the 4D and 10D pictures in this case.  Section
\ref{sec:discussion} contains a discussion of the properties of the
landscape of IIA vacua and compares to other ensembles.  We conclude
in Section 7.  In an Appendix we provide an elementary derivation of the
type IIA Chern-Simons terms in the presence of background fluxes
which are needed for our analysis.

\section{A simple model: $T^6/\bZ_3^2$}
\label{sec:model}

In this section we describe a simple type IIA orientifold
compactification which we will use as an example throughout this
paper.  The model is a $T^6/\bZ_3$ orientifold, modded out by an
additional freely acting $\bZ_3$ symmetry \cite{orbifold,Strominger,bgk}, preserving ${\cal N} =1$ supersymmetry in four dimensions.
A discussion of the stabilization of untwisted moduli for a $T^6/(\bZ_2 \times \bZ_2)$ IIA orientifold appears in \cite{Zwirner}.

This compactification has a fairly small number of moduli and is easy to analyze
explicitly.  There are moduli corresponding to the sizes of the three 2-tori $T^6 = T^2 \times T^2 \times T^2$,  a $B$-field modulus for each $T^2$, and finally the dilaton and a single axion arising
from the 3-form $C_3$; there are no complex structure moduli. Furthermore, there are additional metric and $B$-field moduli associated with blow-ups of 9 singular orbifold points.

In Section \ref{sec:10D}, we show that all these moduli are stabilized in type IIA supergravity when the zero- and three-form fluxes $F_0$ and $H_3$ canceling the tadpole from the orientifold fixed plane are combined with generic four-form fluxes $F_4$.  We demonstrate this by directly calculating the potential for the zero modes.
In Section \ref{sec:4D-applied}, we consider the ${\cal N}=1$ four-dimensional effective supergravity description and show that depending on the signs of the fluxes, these stabilized vacua may be supersymmetric solutions extremizing the flux superpotential.

Let us describe the orientifold in more detail.  We parameterize the torus $T^6$ by three complex
coordinates $z_i = x_i + i y_i$, subject to the periodicity conditions
\begin{equation}
z_i \sim z_i + 1 \sim z_i + \alpha \,,
\label{eq:torus-identification}
\end{equation}
where $\alpha = e^{\pi i/3}$. This torus has a $\bZ_3$ symmetry $T$ under the action
\begin{equation}
T: (z_1, z_2, z_3) \rightarrow
 (\alpha^2 z_1, \alpha^2 z_2, \alpha^2 z_3) \,.
\label{eq:t}
\end{equation}
This transformation has 27 fixed points, and the resulting orbifold is a
singular limit of a Calabi-Yau with Euler character $\chi = 72$.\footnote{Note that while homotopically nontrivial curves
in the original $T^6$ (such as a cycle wrapping once on $x_i$) are not
all trivial after the $\bZ_3$ quotient, such nontrivial
cycles project to elements of the $\bZ_3$ around the fixed points and
are removed when these points are blown up to form a smooth
Calabi-Yau, so that the resulting CY indeed has no $\pi_1$.}
This orbifold was constructed in \cite{orbifold} and
its geometry was analyzed in detail in \cite{Strominger}, where it was also
pointed out that the resulting space has a further $\bZ_3$ symmetry
acting without fixed points according to
\begin{equation}
Q: (z_1, z_2, z_3) \rightarrow
(\alpha^2 z_1 + \frac{1 + \alpha}{3}, \;
\alpha^4 z_2 + \frac{1 + \alpha}{3},\;
 z_3 + \frac{1 + \alpha}{3} ) \,.
\label{eq:q}
\end{equation}
Modding out by this additional $\bZ_3$ leads to a singular limit of a
Calabi-Yau with $\chi = 24$ having 9 $\bZ_3$
singularities.  This
compactification has $h^{2, 1} = 0$ and $h^{1, 1} = 12$, with 9 of the
12 K\"ahler moduli arising from blow-up modes of the 9 singularities.

Following \cite{bgk}, we can construct an orientifold of this $T^6/\bZ_3^2$ orbifold, modding out by ${\cal O} = \Omega_p (-1)^{F_L} \sigma$ where $\Omega_p$ is worldsheet parity, $(-1)^{F_L}$ is left-moving fermion number and $\sigma$ is the reflection
\begin{eqnarray}
\label{Sigma}
\sigma: \; z_i \rightarrow -\bar{z}_i \,,
\end{eqnarray}
for each $i = 1,2,3$.
This gives an ${\cal N}=1$ supersymmetric type IIA orientifold model with an O6 orientifold plane
filling the 4 noncompact directions and wrapping a 3-cycle on the $T^6$.

We are interested in the moduli of this orientifold compactification, corresponding to constant modes
of the various supergravity fields that survive the orbifold and orientifold projections.
Let us begin by discussing the metric on the $T^6$. Invariance of the metric under the action (\ref{eq:q}) of $Q$ dictates that
$g_{ij} = g_{\bar{i} \bar{j}} = g_{i \bar{j}} = 0$ if $i \neq j$ for $i,j = 1,2,3$.
Further, from the invariance of the metric under the action (\ref{eq:t}) of
$T$, it follows that the metric on each $T^2$ is diagonal.  Thus, we
can parameterize the metric on the compact space as
\begin{eqnarray}
\label{Metric}
ds^2 =  \sum_{i=1}^3 \gamma_i \, dz^i d\bar{z}^i =\sum_{i=1}^3 \gamma_i \,  ((dx^i)^2 + (dy^i)^2) \,,
\end{eqnarray}
with 3 real moduli $\gamma_i$ corresponding to the size of each of 3
$T^2$s; all other metric degrees of freedom, including all complex
structure moduli, are projected out.

Consider now the invariant moduli associated with $p$-forms on the compact space.  We can determine how the various modes of each $p$-form field transform under the two $\bZ_3$ symmetries by using
\begin{equation}
T:dz_i \rightarrow \alpha^2 dz_i, \;\;\;\;\;
Q:dz_i \rightarrow \alpha^{2i} dz_i \,.
\end{equation}
The only two-forms which are invariant under both $T$ and $Q$ are
$dz^i \wedge d \bar{z}^i$, which we use to construct a basis $\{w_i\}$, odd under the reflection $\sigma$ (\ref{Sigma}),
\begin{eqnarray}
w_i \,=\, (\kappa\,\sqrt{3})^{1/3} i dz^i \wedge d\bar{z}^i \,, \quad \quad
\int_{T^6/\bZ_3^2}\, w_1 \wedge w_2 \wedge w_3 \,\equiv\, \kappa \,,
\end{eqnarray}
where we have left the overall normalization of the triple intersection $\kappa$ arbitrary.  For later convenience we define a dual basis of even four-cycles $\{\tilde{w}^i\}$,
\begin{eqnarray}
\label{FourCycles}
\tilde{w}^i \,=\, \left({3 \over \kappa}\right)^{1/3}
( i dz^j \wedge d\bar{z}^j) \wedge (i dz^k \wedge d\bar{z}^k) \,, \quad \quad
\int_{T^6/\bZ_3^2}\, w_i \wedge \tilde{w}^j \,=\, \delta_i^j \,,
\end{eqnarray}
where $j$ and $k$ are the two values of $1,2,3$ besides $i$.

The NSNS 2-form potential $B_2$ is odd under the world-sheet orientifold transformation $\Omega_p (-1)^{F_L}$; hence one may have nonzero
\begin{eqnarray}
\label{TwoForm}
B_2 = \sum_{i=1}^3 b_i \, w^i \,.
\end{eqnarray}
These real $b_i$ combine with multiples of the $\gamma_i$,
\begin{equation}
v_i \,\equiv\, \frac{1}{2}\,\frac{1}{(\kappa\,\sqrt{3})^{1/3}}\;\gamma_i\,.
\label{VDef}
\end{equation}
into three complex parameters which will be identified with the K\"ahler moduli of the four-dimensional supergravity studied in Sections \ref{sec:4D} and \ref{sec:4D-applied}.

Because $H^1$ of the resolved orientifold is trivial, there are no moduli
associated with the R-R one-form $C_1$. There is a single modulus
associated with the dilaton $\phi$, as well as its partner, an axion field $\xi$ coming from the RR potential $C_3$, as we now describe.

The three-forms which are invariant under $T$ and $Q$ are the holomorphic 3-form
\begin{equation}
\Omega \,=\, 3^{1/4}\,i\,dz_1\wedge dz_2 \wedge dz_3 \,,
\end{equation}
and its complex conjugate $\bar{\Omega}$. The normalization is fixed
to satisfy the following convenient condition
\begin{equation}
i\,\int_{T^6/\bZ_3^2}\,\Omega \wedge \bar{\Omega} \,=\, 1
\end{equation}
where we used  $i\,\int_{T^2}\, dz_i \wedge d\bar{z}_i \,=\, \sqrt{3}$.

We can decompose $\Omega$ into real and imaginary components
\begin{equation}
\label{OmegaAlphaBeta}
\Omega \,=\, \frac{1}{ \sqrt{2}} \,
\left(\alpha_0 + i\,\beta_0 \right)
\end{equation}
where because under (\ref{Sigma}) we have $\sigma: \Omega \rightarrow \bar{\Omega}$, $\alpha_0$ and $\beta_0$ are even and odd respectively
under orientifold reflection; the orientifold is hence wrapped on the $\alpha_0$ cycle.  They form a symplectic basis,
\begin{equation}
\int_{T^6/\bZ_3^2}\, \alpha_0 \wedge \beta_0 \,=\, 1 \,.
\end{equation}
Under the world-sheet orientifold transformation $\Omega_p (-1)^{F_L}$, $C_{(3)}$ is even, and hence the single modulus $\xi$ of the R-R three-form is
\begin{equation}
\label{ThreeForm}
C_{(3)} \,=\, \xi \,\alpha_0 \,.
\end{equation}
The axion $\xi$ and the dilaton $\phi$ combine into the complex axiodilaton modulus.

In addition to the 4 complex
moduli we have already described,
9 further (complex) K\"ahler moduli are associated with
the blow-ups of the 9 singular points of the orientifold.  Locally,
each blow-up looks like a resolution of $\bC^3/\bZ_3$, and is
parameterized by a scale modulus and a corresponding $B$-field
modulus.  Globally, these moduli can be described in terms of the
metric and $B$-field degrees of freedom on a smooth Calabi-Yau whose
singular limit is the $T^6/\bZ_3$ orientifold.

Although we do not
have an explicit form for the metric on the smooth Calabi-Yau, we can
give a local analysis of these blow-up modes from the point of view
of 10D supergravity, which we do
in Section \ref{sec:10D}.  Furthermore, in the 4-dimensional picture,
the prepotential for these modes is known to leading order, allowing us
to find solutions with all blow-up moduli  stabilized; this analysis is
carried out in Section \ref{sec:4D-applied}.

\section{Moduli stabilization of $T^6/\bZ_3^2$
in classical IIA supergravity}
\label{sec:10D}

We will now directly calculate from the massive type IIA supergravity
action the potential for the moduli of the orientifold
compactification presented in the previous section.  In subsection
\ref{sec:10D-SUGRA} we describe the supergravity action on the
orientifold in the presence of fluxes.  Subsection \ref{sec:tadpole}
solves the equation of motion for the R-R seven-form field $C_{(7)}$
which fixes the tadpole cancellation condition.  In subsection
\ref{sec:moduli} we stabilize the bulk moduli of the compactification
by solving the remaining supergravity equations of motion.  Subsection
\ref{sec:stability} examines potential tachyonic directions, showing
that although for some signs of fluxes there are tachyons, their masses
do not exceed the Breitenlohner-Freedman bound, so that they do not represent
true instabilities; the analysis of section \ref{sec:4D-applied} will show that
only the vacua associated to certain choices of fluxes, all of which have no tachyons,
are supersymmetric.  Finally, subsection
\ref{sec:blow-up} contains a description of the stabilization of the
blow-up modes.

\subsection{Fluxes and the IIA supergravity action}
\label{sec:10D-SUGRA}

In order to stabilize all moduli, we will turn on background fluxes
on the orientifold.  In addition, the orientifold produces a tadpole for the $C_7$ potential,
which must be canceled either by wrapped D6-branes or fluxes; in the
next section we will show how to satisfy the tadpole constraints with fluxes alone.

We will  turn on a constant $F_0$, as well as NS-NS three-form flux $H_3$
and R-R four-form flux $F_4$.   The first two are necessary to cancel the
tadpole, and then the last completes the flux stabilization.  For simplicity
we leave $F_2 = 0$; we discuss the generalization to nonzero $F_2$ in Sections
\ref{sec:4D} and \ref{sec:4D-applied}, and find that most choices of $F_2$
are physically redundant under gauge transformations, while the few physically
inequivalent vacua with nonzero $F_2$ have qualitatively identical behavior
 to the $F_2  = 0$ case we consider here.  $F_6$ only comes into stabilizing
  the axion $\xi$.

Since $B_2$ is odd under the orientifold action, the three-form background $H_3^{\rm bg}$ must multiply the unique odd 3-form (\ref{OmegaAlphaBeta}),
\begin{equation}
\label{HFlux}
H_3^{\rm bg} \,=\, -p \, \beta_0 \,,
\end{equation}
while the four-form flux $F_4$ is expanded in the basis (\ref{FourCycles}) of even 4-cycles,
\begin{equation}
\label{FFlux}
F_4^{\rm bg} \,=\, e_i \, \tilde{w}^i\,.
\end{equation}
We can also turn on four-form flux through 4-cycles associated with the
blow-up modes, as we discuss in subsection \ref{sec:blow-up}.

The presence of nonzero $F_0$ means that instead of ordinary type IIA supergravity, we must use the massive type IIA theory \cite{Romans}, with mass parameter $m_0 = F_0$.  The string frame action is then
\begin{equation}
S = S_{{\rm kinetic}} + S_{\rm CS} + S_{O6} \,,
\label{eq:action}
\end{equation}
where the action is decomposed into a Chern-Simons piece $S_{\rm CS}$,
a piece from the orientifold $S_{{\rm O6}}$, and a ``kinetic'' piece
(everything else).  The kinetic terms are\footnote{We follow the conventions of
\cite{Grimm-Louis-IIA} for the RR fields (including $m_0$) so we can more easily match the 4D superpotential analysis; they are related to those of
Polchinski \cite{Polchinski} by $C_{RR} = C_{RR}^{Polch} / \sqrt{2}$.}
\begin{equation}
S_{\rm kinetic}
 \,=\, \frac{1}{2\kappa_{10}^2} \int d^{10}x\, \sqrt{-g}
\left( e^{-2\phi}(R + 4(\partial_{\mu}\phi)^2 - \frac12 |H^{\rm total}_3|^2)
- (|\tilde{F}_2|^2 + |\tilde{F}_4|^2 + m_0^2)
\right)\,,
\label{eq:kinetic}
\end{equation}
where $2 \kappa_{10}^2 = (2 \pi)^7 {\alpha'}^4$, with field strengths given by
\begin{eqnarray}
\nonumber
H^{\rm total}_3 & = &  dB_2+ H_3^{\rm bg} \,,\\
\tilde{F}_2 & = & dC_1  \,+\, m_0 B_2 \,, \\
\tilde{F}_4& = &dC_3 + F_4^{\rm bg} \,-\, C_1 \wedge H_3 \,-\, \frac{m_0}{2}\, B_2\wedge B_2 \,,
\nonumber
\end{eqnarray}
and $|F_p|^2 = F_{\mu_1 \ldots \mu_p} F^{\mu_1 \ldots \mu_p} / p!$.
We denote by $B_2, C_3$ only the fluctuation part of the form field
around the given background flux.
The Chern-Simons piece takes the form
\begin{eqnarray}
S_{\rm CS} & = &  - \frac{1}{2\kappa_{10}^2}\int \left[
B_2 \wedge dC_3 \wedge dC_3
+ 2B_2 \wedge dC_3 \wedge F_4^{\rm bg} + C_3 \wedge H_3^{\rm bg} \wedge dC_3 \right. \label{eq:CS}\\
 &  & \hspace*{0.7in} \left.
-\frac{m_0}{3}  B_2 \wedge B_2 \wedge B_2 \wedge dC_3
+ \frac{m_0^2}{ 20} B_2 \wedge B_2 \wedge
B_2 \wedge B_2 \wedge B_2 \right]\,.  \nonumber
\end{eqnarray}
The separation of the usual $\int B_2 \wedge F_4 \wedge F_4$
Chern-Simons term into several pieces is needed because topological
fluxes must appear in the field strengths, and in the presence of
fluxes the second and third terms in (\ref{eq:CS}) are not related by
the usual integration by parts.  An elementary derivation of the relevant terms from M-theory is given in the Appendix.  In principle there should be similar contributions
involving the background fluxes in the massive IIA theory of the form
$m_0 B^3 F_4^{\rm bg}$ and $m_0 B^2 H_3^{\rm bg}C_3$; we do not need
such terms for the analysis here.   Quantum type IIA string theory involves a number of subtleties related to the K-theoretic classification of branes and fluxes \cite{KTheory}, some of which generalize the Chern-Simons terms \cite{CS}; these subtleties do not affect our results.

Finally, the contribution of the orientifold fixed plane to the action is given by
\begin{equation}
S_{\rm O6}
\, =\, 2 \mu_6 \,\int_{O6} {d}^7\xi
e^{-\phi}\,\sqrt{-g} - 2 \sqrt{2} \mu_6 \int C_{(7)} \,,
\label{eq:so6}
\end{equation}
where $\mu_p = (2 \pi)^{-p} {\alpha'}^{-(p+1)/2}$ is the D$p$-brane charge and tension, and we have taken into account that the charge of an O$p$-plane is $-2^{p-5}$ that of a D$p$-brane.

Before proceeding to evaluate the $C_7$ tadpole, we remark on the
quantization of the fluxes.  For a canonically normalized $F_p$ field strength, the
usual (cohomological) quantization condition is
\begin{eqnarray}
\label{UsualQuant}
\int F_p =  2 \kappa_{10}^2 \, \mu_{8-p} \, f_p =
  (2 \pi)^{p-1} {\alpha'}^{(p-1)/2} f_p \,,
\end{eqnarray}
with $f_p$ an integer; in our convention the expression for RR fields must be rescaled by a factor of $\sqrt{2}$.  Hence we can write the fluxes we are using in terms of integers $f_0, h_3, f_4^i$ as
\begin{eqnarray}
\label{Quantization}
m_0 = {f_0 \over 2 \sqrt{2} \pi \sqrt{\alpha'}} \,, \quad \quad p =
(2\pi)^2 \alpha' \, h_3 \,, \quad \quad e_i =
\frac{\kappa^{1/3}}{ \sqrt{2}} \, (2 \pi \sqrt{\alpha'})^3\, f_4^i \,.
\end{eqnarray}
The K-theoretic classification of fluxes \cite{KTheory} modifies the condition (\ref{UsualQuant}) in certain circumstances.  The effect potentially relevant to our analysis is that when the first Pontryagin class divided by two $p_1/2$ of the tangent bundle of the compactification manifold is odd,
the $f_4^i$ are half-integers instead of integers \cite{WittenM}; however this shift turns out not to affect any of the cycles in our $T^6/\bZ_3^3$ example, as $p_1$ is always divisible by four for a Calabi-Yau threefold.\footnote{We thank Paul Aspinwall for comments on characteristic classes for Calabi-Yaus.}

\subsection{Cancelling the tadpole}
\label{sec:tadpole}

As is evident from (\ref{eq:so6}), the O6 plane generates a
tadpole for the $C_7$-potential Hodge dual to $C_1$.  This can be
cancelled by adding 2 D6-branes for each O6, but instead we cancel
it using the background fluxes.

One may show by analyzing the RR equations of motion and Bianchi identities, as well as various gauge invariances in the brane actions, that
\begin{eqnarray}
\tilde{F}_6 &\equiv& * \tilde{F}_4 = dC_5 - C_3 \wedge H_3 + {m_0 \over 6} B_2 \wedge B_2 \wedge B_2 \,, \\
\tilde{F}_8 &\equiv& * \tilde{F}_2 = d C_7 - C_5 \wedge H_3 - {m_0 \over 24} B_2 \wedge B_2 \wedge B_2 \wedge B_2 \,.
\end{eqnarray}
The equation of motion for $C_7$ then receives contributions from the $|\tilde{F}_2|^2$ term in (\ref{eq:kinetic}), as well as from the O6-plane in (\ref{eq:so6}).  Integrating over the $\beta_0$ cycle, one finds
\begin{eqnarray}
\int d \tilde{F_2} =  2 \sqrt{2} \kappa_{10}^2\, \mu_6  \,,
\end{eqnarray}
which using $d \tilde{F}_2 = m_0 H_3$ gives the tadpole condition
\begin{eqnarray}
\label{Tadpole}
m_0 p = -  2 \sqrt{2} \kappa_{10}^2 \, \mu_6 = -2 (\sqrt{2} \pi
\sqrt{\alpha'})\,.
\end{eqnarray}
Hence we learn that $m_0$ and $p$ must be of opposite sign.  We note that the quantization condition (\ref{Quantization}) requires
\begin{eqnarray}
m_0 p = (\sqrt{2} \pi \sqrt{\alpha'}) f_0 h_3 \,
\end{eqnarray}
with $f_0, h_3$ integers, so that the minimal charge that can be obtained from $F_0 H_3$ is just that of a single D6-brane.  To satisfy the tadpole (\ref{Tadpole}) we have a very limited set of possibilities for the fluxes: $f_0 h_3 = -2 \rightarrow \pm (f_0, h_3) = (-1,2), (-2,1)$.  Hence there is very little freedom to tune in the $H_3$, $F_0$ sector; what freedom we have will come from $F_4$.

We note that in other models with more 3-cycles, there will be in general $h^{2,1}+1$ tadpole conditions to satisfy; this is to be contrasted with the single $C_4$ tadpole familiar in IIB flux compactifications.  Thus once $F_0$ is nonzero, every mode of $H_3$ will be constrained by a tadpole condition.

\subsection{Stabilizing bulk moduli}
\label{sec:moduli}

Having chosen the $H_3$ and $F_0$ fluxes so as to satisfy the tadpole
cancellation condition (\ref{Tadpole}), we now turn to evaluating the
potential for the moduli and solving the resulting equations of
motion.  We insert the background fluxes (\ref{HFlux}),
(\ref{FFlux}) into the supergravity action,
and write the metric, $B$ field and 3-form field $C_3$ in terms of the
bulk moduli using (\ref{Metric}), (\ref{TwoForm}),
(\ref{ThreeForm});  to determine the potential we assume the modes
$\gamma^i, b^i, \phi, \xi$ are coordinate-independent.

A more complete analysis would include the warp factors in the metric
and the full dependence of the supergravity fields on the compact
directions.  We leave the details of such an analysis for future work;
as we shall see, the model we are considering here admits
solutions in a regime of large volume and weak coupling where these
effects are unimportant.

We begin by considering the RR 3-form field $C_3$.  The single
modulus $\xi$ of this field appears only in the Chern-Simons
term $C_3 \wedge H_3^{bg} \wedge dC_3$, and we note that given the value (\ref{HFlux}) for
$H_3^{bg}$, this term is only nonzero if the remaining $dC_3$ is polarized along the spacetime directions; hence treating this latter mode is necessary for determining the equation for the axion $\xi$.  This field has no physical degrees of freedom; we shall call it $dC_3|_{4D} \equiv {\cal F}_0$ and treat it as a Lagrange multiplier.  A more careful, quantum-mechanical treatment leading to the same result is described in \cite{Beasley-Witten}.

For a field with couplings of the form
\begin{eqnarray}
\label{Lagrange}
S = - {1 \over 2 \kappa_{10}^2} \int \left( {\cal F}_0 \wedge * {\cal F}_0 + 2 {\cal F}_0 \wedge X \right) \,,
\end{eqnarray}
the equation of motion for ${\cal F}_0$ merely sets ${\cal F}_0 = * X$; substituting this back into the action, (\ref{Lagrange}) becomes
\begin{eqnarray}
\label{LagrangeDone}
S = - {1 \over 2 \kappa_{10}^2} \int X \wedge *X \,.
\end{eqnarray}
Hence minimization of these terms in the potential simply sets $X=0$.  Calculating $X$ for the case at hand and integrating over the compact space, we have
\begin{eqnarray}
\label{XEqn}
\int X = 0 = \int \left( F_6^{\rm bg} + B_2 \wedge F_4^{\rm bg} + C_3 \wedge H_3^{\rm bg} - {m_0 \over 6} B_2 \wedge B_2 \wedge B_2 \right) \,,
\end{eqnarray}
which evaluates to an equation for the 3-form axion $\xi$,
\begin{eqnarray}
p  \, \xi = e_0 + e_i b_i - \kappa \, m_0 \, b_1 b_2 b_3 \,,
\end{eqnarray}
where we put $e_0 = \int F_6^{\rm bg}$.

We now solve the equation of motion for the $B$ field components.
Since there are no zero modes of $C_1$ and we have taken $F_2^{\rm bg}
= 0$, the $|\tilde{F}_2|^2$ and $|\tilde{F}_4|^2$ terms are at least
quadratic in $b_i$; the Chern-Simons terms have already been accounted
for in the minimization of $X$.  Thus (\ref{eq:action}) is at least
quadratic in $b^i$, meaning we can consistently find a
solution with $b_i = 0$.

Notice that the term $|\tilde{F}_4|^2$
gives rise to an off-diagonal quadratic term for the $B$-field moduli
of the form $(F^{\rm bg}_4)_{abcd} \, B^{ab}B^{cd}$.  Such a term can lead
to an unstable $B$ mode.  After solving for the rest of the moduli we
return to this term in subsection \ref{sec:stability} and check to see when the quadratic form for the
$B$ moduli is positive definite around the solution.

The moduli that remain are the sizes $\gamma_i$ of the 2-tori and the dilaton $\phi$; we now write the four-dimensional effective potential for these.
We note first that to properly normalize the four-dimensional Einstein term, we pass to a 4D Einstein frame with the redefinition
\begin{eqnarray}
\label{eq:Weyl}
g_{\mu\nu} = {e^{2 \phi} \over \vol} \, g^E_{\mu\nu} \,,
\end{eqnarray}
for the four-dimensional metric only.  We then define the effective potential $V$,
\begin{eqnarray}
S = {1 \over \kappa_{10}^2} \int d^4x \sqrt{-g_E} (-V) \,,
\end{eqnarray}
and find the result
\begin{equation}
V \,=\,
\frac{p^2}{4}\, \frac{e^{2\phi}}{{\rm vol}^2}+ \frac12 \,(\sum_{i = 1}^{3}e_i^2\,v_i^2) \frac{e^{4 \phi}}{{\rm vol}^{3}} + \frac{m_0^2}{2}\, \frac{e^{4\phi}}{\rm vol}  - \sqrt{2}\,|m_0\, p|\, \frac{e^{3\phi}}{{\rm vol}^{3/2}} \,,
\label{eq:potential}
\end{equation}
where the four terms are from the $|H_3|^2$, $|\tilde{F}_4|^2$ and $m_0^2$
terms in (\ref{eq:kinetic}) and the O6 Born-Infeld piece in (\ref{eq:so6}),
respectively; the $|\tilde{F}_2|^2$
and O6 Chern-Simons terms cancel according to the tadpole cancelation
condition (\ref{Tadpole}).
We have defined the volume of compactification
\begin{equation}
{\rm vol} \,\equiv\, \int_{T^6/\bZ_3^2}\sqrt{g_6}
\,=\, \frac{1}{8\,\sqrt{3}}\, \gamma_1 \gamma_2 \gamma_3  \,\equiv\, \kappa\,v_1 v_2 v_3 \,,
\label{eq:volume}
\end{equation}
and written (\ref{eq:potential}) in terms of the rescaled metric components $v_i$ (\ref{VDef}).
The evaluation of the O6-plane contribution to the potential (\ref{eq:potential}),
\begin{eqnarray}
V_{O6} =  - 2 \kappa_{10}^2 \, \mu_6 \, {e^{3\phi} \over \vol^2} \int d^3x \sqrt{g_3} \,,
\end{eqnarray}
was carried out using the calibration formula \cite{BBS} for special Lagrangian 3-cycles, which for us reads
\begin{eqnarray}
\int d^3x \sqrt{g_3} = 2 \sqrt{2}\, \vol^{1/2}\,  \int \Re \Omega = 2 \, \vol^{1/2} \, \int \alpha_0 \,.
\end{eqnarray}
We now want to solve the equations
\begin{equation}
\frac{\partial V}{ \partial \phi}  = \frac{\partial V}{\partial v_i} = 0 \,.
\label{eq:tosolve}
\end{equation}
The structure of the $\partial_{v_i}$ equations is
\begin{equation}
\frac{ F({\rm vol}, \phi)}{v_i}  + e_i^2\, v_i\, G ({\rm vol}, \phi) = 0,
\label{eq:solvable}
\end{equation}
where $F, G$ are some functions of ${\rm vol}$ and $\phi$.  Thus, we can reduce to
two degrees of freedom using $v_i = v/|e_i|$, giving the
simplified potential
\begin{equation}
V (D,  v) \,=\,
\frac{m_0^2}{2 E}\,e^{4 D}\,v^3 - \sqrt{2}\, |m_0\, p |\, e^{3D}
+ \frac{p^2}{4}\, \frac{e^{2D}}{v^3}\,E
+ \frac32\,\frac{e^{4D}}{v}\,E  \,,
\end{equation}
where $E = |e_1 e_2 e_3 |/\kappa$ (${\rm vol} \,=\, v^3/E$)
and we have also introduced the 4-dimensional dilaton
\begin{equation}
\label{4ddil}
e^{D} \,=\, \frac{e^{\phi}}{\vol^{1/2}} \,.
\end{equation}
Rescaling $e^{D} \,=\, |p |\, \sqrt{|m_0 |/E}\, g$ and
$v \,=\, \sqrt{E/|m_0 |}\, r^{2}$, the potential becomes
\begin{equation}
\frac{1}{ \lambda} V(g, r) \,=\,  \frac12\,g^4\,r^6  - \sqrt{2}\,g^3
+ \frac14\,\frac{g^2}{ r^{6}}  + \frac32\,\frac{g^4}{r^{2}}
\label{eq:vxy}
\end{equation}
where $\lambda \,=\, p^4\,|m_0|^{5/2}\,E^{-3/2}$.

Now, we proceed to find the extremum of (\ref{eq:vxy}).  We have
\begin{equation}
g \partial_g V + 2r \partial_rV
= \lambda  g^4r^6\left[ 4-{3 \over \sqrt{2}} \left( \frac{ 1}{gr^6}  \right)
- \frac{5}{4}  \left( \frac{1}{ gr^6}  \right)^2 \right] = 0,
\end{equation}
which implies
$gr^6  =  {5}/ ({4 \sqrt{2}})$. Plugging in $ g = 5/(4 \sqrt{2} r^6)$ into $\partial_gV = 0$ gives
$r^8 =   {25}/{9}$. We thus have the solution
\begin{eqnarray}
\label{eq:TenDSoln}
v_i & = &  \frac{v}{|e_i |} \,=\,
\frac{1}{|e_i |}\,\sqrt{\frac53\left|\frac{e_1e_2e_3}{\kappa\,m_0} \right|}\,,  \\
e^{D} & = &  |p| \,\sqrt{ \frac{27}{160}\left|\frac{\kappa\,m_0}{e_1e_2e_3 }\right|}\,,\nonumber
\end{eqnarray}
or equivalently in terms of the 10D metric and dilaton,
\begin{eqnarray}
 ds^2  & = &
\left(\frac{1}{9 \kappa}  \right)^{1/6} \;
\sqrt{5 \left | \frac{e_1 e_2e_3}{m_0} \right | }
\;\sum_{i = 1}^{3}  \frac{1}{ | e_i |}
\, dz^id \bar{z}^i \,, \\
e^{\phi} & = & \frac{3}{4}\,|p|\,  \left( \frac{5}{12}
\frac{\kappa}{| m_0 e_1 e_2e_3 | }
\right)^{1/4}  \,.
\end{eqnarray}
Note that the $\kappa$ dependence cancels out when the $e_i$ are expressed in
terms of the quantized fluxes (\ref{Quantization}).

One can show that
\begin{equation}
6g\partial_gV  -r\partial_r V
= 18V + 12\lambda\,\frac{g^4}{r^2}  \,.
\end{equation}
Thus, for the solutions (\ref{eq:TenDSoln}) satisfying $\partial_g V = \partial_r V  =0$, the energy $V$ is always negative:
\begin{eqnarray}
\label{PotentialSoln}
V = - {2 E \over 3 v} \, e^{4D} \,,
\end{eqnarray}
and the 4D space-time is anti-de Sitter.

\begin{figure}
\begin{center}
\epsfig{file=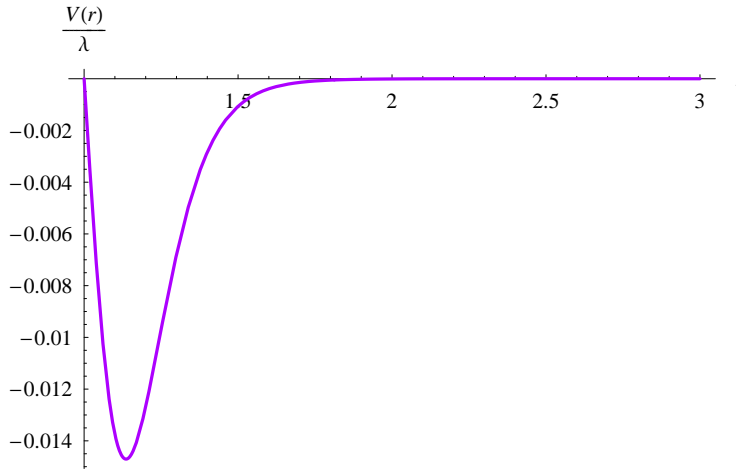,width=10cm}
\end{center}
\caption{\label{plot1} The potential  $\frac{1}{\lambda}\,V(r, g)$
on solutions for $g$ as a function of $r$.}
\end{figure}

\begin{figure}
\begin{center}
\epsfig{file=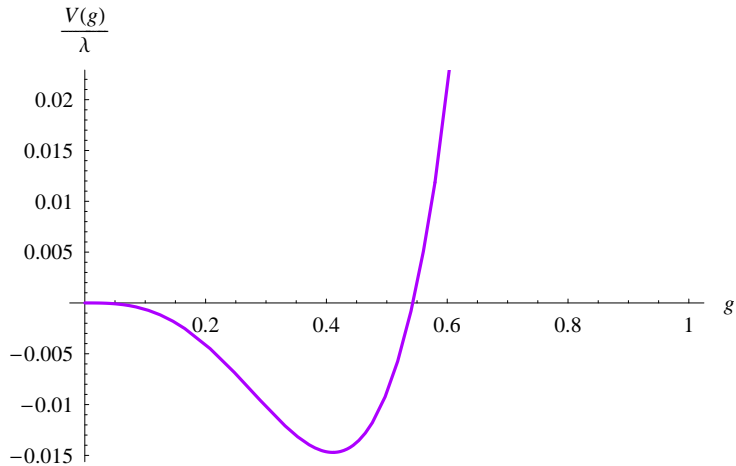,width=10cm}
\end{center}
\caption{\label{plot2} The potential  $\frac{1}{\lambda}\,V(r, g)$
on solutions for $r$ as a function of $g$.}
\end{figure}

The solutions (\ref{eq:TenDSoln}) stabilize all moduli for any choice of $m_0, p$
satisfying the tadpole condition (\ref{Tadpole}) and any four-form fluxes $e_i$.
Because the four-form flux parameters $e_i$ are not constrained by the tadpole,
we have an infinite family of IIA vacua with this orientifold compactification.

The shape of the potential and AdS minima
are exhibited in figures \ref{plot1} and \ref{plot2}.
Note that the finite distance minima for dimensionless
variables $r$ and $g$ will correspond to minima at parametrically large radius
and small coupling in terms of dimensionful parameters.

Scaling all the $e_i$ as $e_i \sim \bar{e}$, we find that the metric
components $\gamma_i$ scale as $\bar{e}^{1/2}$ and hence the volume
goes as $\bar{e}^{3/2}$, while the string coupling $e^{\phi} \sim
\bar{e}^{-3/4}$ and the vacuum energy goes as $-\bar{e}^{-9/2}$.
Thus, the infinite family of compactifications has parametrically
increasing volume and decreasing string coupling.  As we will discuss
further in section \ref{sec:validity}, the solutions are effectively
four-dimensional at low energies, unlike the familiar Freund-Rubin
models, which also arise in infinite families.  This is (granting the
controlled stabilization of the blow-up modes, which we discuss in
\S3.5) the primary result of this paper: a class of four-dimensional
vacua with all moduli stabilized by fluxes in a controlled regime
where corrections can be made arbitrarily small.

\subsection{Stability analysis}
\label{sec:stability}

Because the $e_i$ appear quadratically in the potential (\ref{eq:potential}),
the solution (\ref{eq:TenDSoln}) exists for any choice of sign on the four-form
fluxes; this is manifested by the absolute values in the solution.  The sign of $m_0$ is also arbitrary, although from (\ref{Tadpole}) we must have $\sgn (m_0 p) < 0$.

As we shall see in Section \ref{sec:4D-applied}, not all choices of sign
for the fluxes lead to supersymmetric vacua at large volume.  This suggests
that some of the solutions (\ref{eq:TenDSoln}) could have instabilities.
We now consider the quadratic form for the fields $B_2$ and $C_3$ around the solutions
(\ref{eq:TenDSoln}) and look for possible tachyonic modes.

The $B_2$ field appears in the
$|\hat{F}_2|^2$ and $|\tilde{F}_4|^2$ terms of (\ref{eq:kinetic}).  In the
background given by the solution (\ref{eq:TenDSoln}), these terms
give contributions quadratic in $B_2$ of the form
\begin{equation}
- {1 \over 2 \kappa_{10}^2} \int d^{10}x \sqrt{-g} \, ( |\hat{F}_2|^2 + |\tilde{F}_4^2| ) \,\rightarrow\, -{1 \over 2 \kappa_{10}^2} \int (
 m_0^2\,B_2 \wedge * B_2 -m_0\, B_2 \wedge B_2 \wedge * F^{\rm bg}_4) \,,
\end{equation}
while from eliminating the Lagrange multiplier ${\cal F}_0$, we derive a mixing of $B_2$ with $C_3$ fluctuations (\ref{LagrangeDone}), (\ref{XEqn}).  Hence we also need the kinetic terms for both,
\begin{equation}
- {1 \over 2 \kappa_{10}^2} \int d^{10}x \sqrt{-g} \, ( {1 \over 2} e^{-2 \phi} |H_3|^2 + |\tilde{F}_4|^2) \, \rightarrow \, - {1 \over 2 \kappa_{10}^2} \int \,( {1 \over 2} e^{-2 \phi} \, dB_2 \wedge * dB_2 + dC_3 \wedge * dC_3)  \,.
\end{equation}
These expressions lead to the quadratic action for $b_i$ and $\xi$ fluctuations around the background (\ref{eq:TenDSoln}),
\begin{eqnarray}
S_{axion} &=& {1 \over 2 \kappa_{10}^2} \int d^4x \sqrt{-g_E} \, \left(
\sum_{i = 1}^{3}  \left[ -\frac12 \partial_\mu \tilde{b}_i \partial^\mu \tilde{b}_i  - e^{4D}( m_0^2 \,{\rm vol}\, \tilde{b}_i^2 -2\, m_0 \, \tilde{b}_1 \tilde{b}_2 \tilde{b}_3\,
\frac{e_i v_i}{\tilde{b}_i})\right] \right. \hspace*{0.4in}\\
& &
\hspace*{1.5in}
\left.-\frac12 \partial_\mu x \partial^\mu x  - {e^{4D} \over\vol}
(\tilde{b}_1 e_1 v_1 + \tilde{b}_2 e_2 v_2 + \tilde{b}_3 e_3 v_3  - {p
  \over \sqrt{2}} e^{-D} x )^2  \right)\,, \nonumber
\end{eqnarray}
where we have normalized the kinetic terms by defining $\tilde{b}_i \equiv b_i/v_i$, $x \equiv \sqrt{2} e^D \xi$.
The mass-squared matrix for the coupled $\tilde{b}_i$, $x$ sector is then
\begin{equation}
M^2_{ij} = 2 \,|m_0| \, e^{4D} \, v\left(
\begin{array}{cccc}
 \frac{34}{15} & \frac35 s_1 s_2 - s_3  & \frac35 s_1 s_3 - s_2  & \frac45 s_1  \\
 \frac35 s_1 s_2- s_3 &  \frac{34}{15} & \frac35 s_2 s_3 - s_1 & \frac45 s_2 \\
\frac35 s_1 s_3 - s_2 & \frac35 s_2 s_3  - s_1   & \frac{34}{15} & \frac45 s_3 \\
\frac45 s_1 & \frac45 s_2 & \frac45 s_3 & \frac{16}{15}
\end{array} \right) \,,
\end{equation}
where $s_i \equiv \sgn (m_0 e_i) = \pm 1$.  It is easy to check that when
\begin{equation}
\label{SecThreeSigns}
e_1 e_2 e_3 m_0 \,<\, 0 \,,
\end{equation}
the matrix $M^2_{ij}$ is positive definite, whereas when $e_1 e_2 e_3 m_0 \,>\,0$, $M^2_{ij}$ has a negative eigenvalue, $M^2_{tachyon} = - (2/15) ( 2\, |m_0| \,e^{4D}\, v)$.

Because the vacuum solutions are in anti-de Sitter space, it is not
enough to find that a tachyonic mode exists for an instability to be
present; tachyons whose negative mass-squared is above
(less negative than) the Breitenlohner-Freedman bound \cite{BF},
\begin{eqnarray}
m^2 \geq m^2_{BF} \equiv - \frac34 |V| \,,
\end{eqnarray}
do not generate an unstable perturbation.  Using (\ref{PotentialSoln}), we find
\begin{eqnarray}
\label{BFRatio}
{M^2_{tachyon} \over m^2_{BF}} = \frac{8}{9} \,.
\end{eqnarray}
Since this is less than $1$, the tachyon satifies the Breitenlohner-Freedman bound, and does not lead
to an instability.  Notice that this ratio is independent of the magnitudes of any of the fluxes; in general all the quadratic $b$ and $\xi$ fluctuations have their masses set by the $AdS$ scale alone.

In principle, tachyons may also arise in the spectrum of metric or
dilaton fluctuations.  It is fairly straightforward to expand all
metric and dilaton modes around the solution and confirm that the
resulting mass matrix is positive definite.

Thus, all of our solutions (\ref{eq:TenDSoln}) are perturbatively stable.  We shall see in section \ref{sec:4D-applied} that only for certain signs of the fluxes are the solutions supersymmetric.  The possible existence of quantum instabilities in these vacua is an interesting question we leave for the future.

\subsection{Blow-up modes}
\label{sec:blow-up}

Before closing this section, we discuss stabilizing the K\"ahler
moduli associated to blow-up modes of the $T^6/\bZ_3^2$ orbifold.
A complete treatment of these degrees of freedom from ten-dimensional
supergravity is difficult because
we do not know the explicit  form of the metric for the smooth
Calabi-Yau which arises when all the singularities are blown up.  In
this section we simply consider the blow-up modes locally and show
that they can be stabilized by a 4-form flux on the $\bC\bP^2$ cycle
which blows up the local $\bC^3/\bZ_3$ singularity.

The local analysis we carry out here is valid as long as the scale of
the blow-up mode is much smaller than the scale of the
compactification determined by the untwisted modes (while still very
large in string units, so supergravity can be trusted).  As we shall
see, this can be guaranteed by choosing the flux on the $\bP^2$ to be
small compared to the untwisted fluxes $e_i$.  Because we are already
making a local approximation, we drop constant factors of order 1 in
the analysis here and simply find the general form of the stabilized
blow-up modes.
In Section \ref{sec:4D-applied} we consider the
complete set of blow-up modes from the 4D supergravity picture, where
all information needed to find the precise global form of the
supergravity solution is contained in the prepotential.

Consider the noncompact $\bC^3/\bZ_3$ singularity.  The
resolution of this singularity by a $\bP^2$ gives a one-parameter
family of metrics on a line bundle ${\cal O} (-3)$ over $\bP^2$.  An
explicit form of the metric is given by \cite{Ganor-Sonnenschein}
\begin{equation}
ds^2 \,=\,\frac{r^2}{2}\, g^{{\rm FS}}_{i \bar{j}}\, dz_id \bar{z}_j + F(r)^{-1} dr^2 +
\frac{r^2}{9}\,  F (r) (d \theta -3A)^2,
\end{equation}
where $F (r) = 1-a^6/r^6$, $a$ parameterizes the blow-up ($r \ge a$ for
any fixed $a$), $g^{{\rm FS}}_{i
\bar{j}}$ is the Fubini-Study metric on $\bP^2$, and $A$ is a
one-form with $dA = ig^{{\rm FS}}_{i \bar{j}}dz_i \wedge d\bar{z}_j$.  We want
to put an integral flux $f$ on the $\bP^2$ and consider the effect
on the 4D potential when this local blow-up occurs inside a much
larger compact manifold.

The only terms in the 10D supergravity action (\ref{eq:kinetic}) which
are relevant are the $m_0^2$ and $| \tilde{F}_4 |^2$ terms.  As in the case of
the bulk moduli, when $F_2 = 0$ we can consistently set $B_2 \,=\, 0$.
There are potentially tachyons arising from new $F_4 B_2^2$ terms.
There is only a single $F_4$ and a single $B_2$; the corresponding cubic
intersection form on the blow-up cycle is nonzero (as we discuss in
more detail in Section \ref{sec:4D-applied}), so the condition
that the vacuum be tachyon-free fixes the sign of the 4-form flux
allowed.  The $m_0^2$ term will, as in (\ref{eq:potential}), take the
form $m_0^2\,e^{4\phi}/{\rm vol}$ where ${\rm vol}$ is the total volume of the
compactification.

From the form of the metric, we see that at
blow-up parameter $a$ we have roughly removed a region of radius $a$
and volume $a^6$ from the volume ${\rm vol}_0$ of the full compactification
with no blow-up.  More precisely, neglecting the cross-terms $d \theta
\wedge A$, the volume form is $\sqrt{g} \,=\, r^6 \sqrt{g^{\rm FS}}/18$.
Corrections to this volume form are small near $r \sim a$, where
the major deformation away from the singular metric occurs.  Thus, the
correction to the volume is ${\cal O} (a^6)$ and so the volume is ${\rm vol}
\,\sim\, {\rm vol}_0-B\, a^6$ where $B$ is a constant.

We can treat the $| \tilde{F}_4 |^2$ term similarly.  Because (neglecting
backreaction) the four-form flux is on the $\bP^2$, we have $| \tilde{F}_4 |^2
\sim r^{-8}f^2$.  Integrating this over the volume gives $\int_ar^{-3}
\sim 1/a^2$, and using (\ref{eq:Weyl}) we then have a total potential
of the form
\begin{equation}
V_{\rm blow-up} \,\sim\, m_0^2 \frac{e^{4 \phi}}{{\rm vol}}
+ C\,f^2 \frac{e^{4 \phi}}{a^2\,{\rm vol}^2}\,,
\end{equation}
where $C$ is a constant and ${\rm vol} \,\sim\, {\rm vol}_0-B\, a^6$.
The minimum of the potential for $a$ is then (for small $a$)
\begin{equation}
a^8 \,\sim\, \frac{C\,f^2}{B\,m_0^2}  \,.
\end{equation}
Thus, we see that
\begin{equation}
a \,\sim\, \left( \frac{f}{m_0}  \right)^{1/4} \,.
\label{TenDBlowUp}
\end{equation}
We see that as long as $f \,\ll\, \bar{e}$ we have stabilized the blow-up
mode at a scale much smaller than the untwisted moduli parameterizing
the size of the overall compactification.
So working in the regime $m_0 \, \ll\, f \, \ll\, \bar e$, we can accomplish
controlled stabilization of the blow-up modes, in a regime where
the supergravity approximation is valid.
We derive the precise
formula for the stabilized blow-up moduli in Section
\ref{sec:4D-applied} using the four-dimensional approach.

\section{IIA flux vacua in 4D ${\cal N}= 1$ supergravity}
\label{sec:4D}

The orientifold of $T^6/\bZ_3^2$ we have studied so far is a
particular case of the general class of ${\cal N}=1$ supersymmetric
orientifolds of Calabi-Yau compactifications of type IIA string
theory.  The effective theory of these models is an ${\cal N}=1$
four-dimensional supergravity, characterized by a superpotential $W$
generated by the fluxes for the moduli fields surviving the
orientifold projection.

In this section, we review the derivation of the flux superpotential by Grimm and Louis \cite{Grimm-Louis-IIA} (for earlier related work see \cite{Louis}; the form of these superpotentials was proposed in \cite{Fluxpot} and also derived in \cite{Zwirner}), and then analyze the general structure of the supersymmetric vacua corresponding to solutions of the conditions $DW=0$.  The equations for the K\"ahler moduli decouple from the other fields and can be solved separately, as do the equations for the
complex structure moduli; the dilaton is then fixed by an equation involving expectation values for the rest of the fields.

We show that in general, all geometric moduli can be frozen by fluxes; axionic partners of the complex structure moduli arising from $C_3$ remain unfixed, however.  In the next section, we turn this analysis on the example of the $T^6/\bZ_3^2$ orientifold, and find results in agreement with the previous sections.

\subsection{Orientifold projection on ${\cal N}=2$ moduli}

The four-dimensional effective theory of type IIA string theory on a Calabi-Yau threefold
is an ${\cal N}=2$ supergravity.  The moduli space is a product of two factors, one containing the vector multiplets (the K\"ahler moduli) and the other the hypermultiplets (the complex structure moduli and dilaton); the metric on each factor is determined by a K\"ahler potential.  The orientifold projection to an ${\cal N}=1$ theory reduces the size of each moduli space, as we review below.

The orientifold projection ${\cal O} = \Omega_p (-1)^{F_L} \sigma$ is the composition of worldsheet parity $\Omega_p$, left-moving fermion number $(-1)^{F_L}$ and an antiholomorphic involution of the Calabi-Yau $\sigma$.  The involution must act on the K\"ahler form $J$ and holomorphic 3-form $\Omega$ as
\begin{eqnarray}
\label{OrientOmega}
\sigma^* J = -J \,, \quad \sigma^* \Omega = e^{2i\theta} \overline\Omega \,,
\end{eqnarray}
where $\theta$ is some phase.  The fixed loci of $\sigma$ are special Lagrangian three-cycles $\Sigma_n$ satisfying
\begin{eqnarray}
J|_{\Sigma_n} = 0 \,, \quad \Im (e^{-i\theta} \Omega)|_{\Sigma_n} = 0 \,.
\end{eqnarray}
Orientifold six-planes  (O6s) fill spacetime and wrap the $\Sigma_n$.  One may always eliminate $\theta$ by a redefinition of $\Omega$, and we shall do so in the following.

For modes of the massless ten-dimensional fields to be invariant under the orientifiold projection, they must transform under the antiholomorphic involution as
\begin{eqnarray}
\label{OrientAction}
\sigma^* g_{\mu\nu} = g_{\mu\nu} \,, \quad \sigma^* B_2 = - B_2 \,, \quad \sigma^* \phi = \phi \,, \quad \sigma^* C_1= - C_1 \,, \quad \sigma^* C_3 = C_3 \,.
\end{eqnarray}

\subsubsection{K\"ahler moduli space}

Before the orientifold projection, the vector multiplet moduli space is $h^{1,1}$-dimensional, the moduli corresponding to the expansion of the complexified K\"ahler form
\begin{eqnarray}
J_c \equiv B_2 + i J \,,
\end{eqnarray}
in a basis of $(1,1)$-forms.  Under the projection, the space of $(1,1)$-forms $H^{1,1}$ decomposes into even and odd subspaces, $H^{1,1} = H^{1,1}_+ \oplus H^{1,1}_-$, of dimensions $h^{1,1}_-$ and $h^{1,1}_+ = h^{1,1} - h^{1,1}_-$, respectively.
{}From (\ref{OrientAction}) we see that the surviving modes of $J_c$ are associated with odd forms, and hence we find $h^{1,1}_-$ surviving complex moduli $t_a$:
\begin{eqnarray}
J_c = \sum_{a=1}^{h^{1,1}_-} t_a w_a \,, \quad \quad t_a = b_a + i v_a \,,
\end{eqnarray}
with $\{w_a\}$ a basis for $H^{1,1}_-$.

Hence the orientifold reduces the K\"ahler moduli space of the
${\cal N}=2$ theory to a subspace without disturbing the moduli space complex structure.  The K\"ahler potential for the reduced space is
simply inherited from the ${\cal N}=2$ theory:
\begin{eqnarray}
\label{eq:kahlerpot}
K^K(t_a) = - \log ( {4 \over 3} \int J \wedge J \wedge J) = - \log ( {4 \over 3} \kappa_{abc} v_a v_b v_c ) \,,
\end{eqnarray}
where we defined the triple intersection
\begin{eqnarray}
\kappa_{abc} \equiv \int w_a \wedge w_b \wedge w_c \,.
\end{eqnarray}
There are also ${\cal N}=1$ vector multiplets associated to $H^{1,1}_+$ that survive the projection, but these contain no scalars and will not interest us.

\subsubsection{Complex structure moduli space}

Before the projection, the hypermultiplet moduli space is quaternionic.  To define the complex structure moduli, as usual one chooses a basis for harmonic 3-forms $H^3$, $\{\alpha_{\hat{K}}, \beta_{\hat{L}} \}$, where $\hat{K}, \hat{L} = 0 \ldots h^{2,1}$ and $\int \alpha_{\hat{K}} \wedge \beta_{\hat{L}} = \delta_{\hat{K},\hat{L}}$.  One can expand the holomorphic 3-form in this basis,
\begin{eqnarray}
\Omega = Z_{\hat{K}} \alpha_{\hat{K}} - g_{\hat{L}} \beta_{\hat{L}} \,,
\end{eqnarray}
and the complex $Z_{\hat{K}}$ can be taken as
homogeneous
coordinates on the complex structure moduli space; we may call the
inhomogeneous coordinates $z_{K}$, $K = 1 \ldots h^{2,1}$.  The complex
space of the $z_K$ is promoted to a quaternionic space as each $z_K$
is joined by the axionic modes $\xi_K$, $\tilde\xi_K$ defined as
\begin{eqnarray}
C_3 = \xi_{\hat{K}} \alpha_{\hat{K}} - \tilde\xi_{\hat{L}} \beta_{\hat{L}} \,,
\end{eqnarray}
while $\xi_0$ and $\tilde\xi_0$ combine with the dilaton $\phi$ and the dual of $B_2$ polarized along spacetime to form the universal hypermultiplet.  The moduli space is thus $4(h^{2,1} + 1)$-real dimensional.

Under the orientifold, the relevant space of harmonic forms again
decomposes into even and odd subspaces, $H^3 = H^3_+ \oplus H^3_-$,
where each of $H^3_+$ and $H^3_-$ is $h^{2,1}+1$-real dimensional.
The even and odd bases are $\{ \alpha_k, \beta_\lambda \}$ and $\{
\alpha_\lambda, \beta_k \}$, respectively, where $k = 0 \ldots
\tilde{h}$ and $\lambda = \tilde{h}+1 \ldots h^{2,1}$; the parameter
$\tilde{h}$ determining how many $\alpha$s are even is
basis-dependent.  The orientifold condition (\ref{OrientOmega}) with $\theta = 0$
requires
\begin{eqnarray}
\label{OmegaCoord}
\Im Z_k = \Re g_k = \Re Z_\lambda = \Im g_\lambda = 0 \,.
\end{eqnarray}
Two of these conditions are constraints on the moduli, while the other two follow automatically for a space admitting the antiholomorphic involution $\sigma$. We see that for each complex $z_k$, only one real component survives the projection.  The condition (\ref{OrientAction}) that $C_3$ be even also truncates the space of axion fields in half to $\{ \xi_k, \tilde\xi_\lambda \}$.
Consequently for each quaternionic modulus, one complex field
survives: a real or imaginary part of the complex structure modulus
and an axion.

The universal hypermultiplet is also cut in half, as $\phi$ and one of $\xi_0, \tilde\xi_0$ survive.  One can summarize all the surviving moduli in the object
\begin{eqnarray}
\Omega_c \equiv C_3 + 2 i \Re (C \Omega) \,,
\end{eqnarray}
where the ``compensator" $C$ incorporates the dilaton dependence through
\begin{eqnarray}
\label{Compensator}
C \equiv e^{-D + K^{\rm cs}/2} \,, \quad \quad e^D \equiv \sqrt{8} e^{\phi + K^K/2} \,=\,
\frac{e^{\phi}}{\sqrt{\rm vol}} \,.
\end{eqnarray}
Here $e^D$ is the four-dimensional dilaton,
equivalent to the previous definition (\ref{4ddil})
using $\int\, J \wedge J \wedge J \,=\, 6\, {\rm vol}$,
and $K^{\rm cs}$ is the K\"ahler potential for complex
structure moduli restricted to the surviving modes
\begin{eqnarray}
\label{KahlerCS}
K^{\rm cs} = - \log ( i \int \Omega \wedge \overline\Omega) = - \log 2 (\Im Z_\lambda \Re g_\lambda - \Re Z_k \Im g_k ) \,.
\end{eqnarray}
The surviving moduli are then the expansion of $\Omega_c$ in $H^3_+$:
\begin{eqnarray}
N_k &\equiv& {1\over 2} \int \Omega_c \wedge \beta_k = {1 \over 2} \xi_k + i \Re (C Z_k) \,, \\
T_\lambda &\equiv& i \int \Omega_c \wedge \alpha_\lambda = i \tilde\xi_\lambda - 2 \Re (C g_\lambda) \,.
\end{eqnarray}
Note that including the dilaton via $C$ means all $h^{2,1}+1$ complex
modes are physical; the compensator essentially trades the irrelevant scale
factor of $\Omega$ for the physically relevant dilaton field.

Thus in contrast to the K\"ahler case, where $h^{1,1}_-$ complex
moduli are preserved and the rest removed, for the hypermultiplet
moduli space each quaternion is cut in half, leaving always
$h^{2,1}+1$ complex moduli.  How many are $N_k$ and how many are
$T_\lambda$ is basis-dependent; there is always a basis where
$\tilde{h} = h^{2,1}$, and all moduli are $N_k$, leaving the real parts of the complex structure moduli, the $\xi_k$ and the dilaton.

The K\"ahler potential for the surviving fields is
\begin{eqnarray}
K^Q = - 2 \log (2 \int \Re (C \Omega) \wedge * \Re (C \Omega) = 4 D \,,
\end{eqnarray}
where in the last step one used the identity
\begin{eqnarray}
\label{ComplexIdentity}
\int \Re (C \Omega) \wedge * \Re (C \Omega) = \Im (C Z_\lambda) \Re (C
g_\lambda) - \Re (C Z_k) \Im (C g_k) = e^{-2D}/2 \,, \hspace*{0.2in}
\end{eqnarray}
derived using (\ref{KahlerCS}) and the definition (\ref{Compensator}) of $D$.

\subsection{Fluxes and superpotential}

One may turn on nonzero fluxes of the NSNS and RR field strengths consistent with the orientifold projection.  Using (\ref{OrientAction}), we find $H_3$ and $F_2$ must be odd, while $F_4$ is even.  We write the fluxes as
\begin{eqnarray}
H_3 = q_\lambda \alpha_\lambda - p_k \beta_k \,, \quad F_2 = - m_a w_a \,, \quad F_4 = e_a \tilde{w}^a \,, \quad F_0 = m_0 \,,
\end{eqnarray}
where we have used the fact that $H^{2,2}_+$ is the Poincar\'e dual of
$H^{1,1}_-$ since the volume form $J \wedge J \wedge J$ is odd.  The
$F_0$ flux $m_0$ is the mass parameter of massive type IIA
supergravity; an additional parameter $e_0 \,=\, \int F_6$ will arise
as well.

Dimensionally reducing the massive IIA supergravity, neglecting
the backreaction of the fluxes and other local sources,
it was shown by Grimm and Louis in \cite{Grimm-Louis-IIA} that the
resulting potential can be written in the form
\begin{eqnarray}
\label{Potential}
V = e^K \left(\sum_{i,j, = \{ t_a, N_k, T_\lambda \}} K^{ij} D_i W \overline{D_j W} - 3 |W|^2 \right) + m_0 e^{K_Q} \Im W^Q \,,
\end{eqnarray}
where $K = K^K + K^Q$, and where the superpotential $W$ is given by
\begin{eqnarray}
\label{Superpotential}
W(t_a, N_k, T_\lambda) &=& W^Q(N_k, T_\lambda) + W^K(t_a) \,, \\
W^Q(N_k, T_\lambda) &=& \int \Omega_c \wedge H_3 =- 2 p_k N_k - i
q_\lambda T_\lambda \,,\nonumber\\
& = & -p_k \xi_k + q_\lambda \tilde{\xi}_\lambda
+ 2i \left[ -p_k \Re  (CZ_k) + q_\lambda \Re (C g_\lambda) \right]\,,\\[0.1in]
W^K(t_a) &=& e_0 + \int J_c \wedge F_4 - {1 \over 2} \int J_c \wedge
J_c \wedge F_2 - {m_0 \over 6} \int J_c \wedge J_c \wedge J_c \,, \nonumber\\
&=& e_0 + e_a t^a + {1 \over 2} \kappa_{abc} m_a t_b t_c - {m_0 \over 6} \kappa_{abc} t_a t_b t_c \,,
\end{eqnarray}
with $D_i$ the K\"ahler covariant derivative $D_i W \equiv \partial_i W + W
\partial_i K$.  The constant term $e_0$
comes from the space-time dual of $F_4$ polarized in the noncompact
directions, as in Section  \ref{sec:moduli} and as discussed in more
detail in \cite{Beasley-Witten,Grimm-Louis-IIA}, but may equivalently be thought
of as the integrated flux of $F_6$.

When the tadpole conditions are satisfied,
the last term in (\ref{Potential}) cancels with contributions from
local (O6 and D6) sources, and hence is absent in the total
potential.\footnote{This term and the local contributions against
which it cancels were not mentioned in \cite{Grimm-Louis-IIA}.}
Consequently, the potential is completely characterized by the
superpotential $W$ (\ref{Superpotential}).

\subsection{Supersymmetric vacua}

The superpotential (\ref{Superpotential}) was derived by comparing to
the dimensionally reduced ten-dimensional supergravity theory
neglecting backreaction.  For a general background, such an
approximation cannot be used; not only will backreaction complicate
the analysis, but contributions such as worldsheet instanton
corrections that cannot be described in the language of
ten-dimensional supergravity will appear.  Corrections of this
nature can be described naturally in the four-dimensional language;
for example, worldsheet instanton corrections to the K\"ahler
potential are well-known and can in some cases be calculated.  In the
regime of validity of effective field theory, the most useful
description of the system is in terms of the four-dimensional
quantities $W$, $K$, which one may use to attempt to determine the
vacua and dynamics in terms of the properly corrected superpotential
and K\"ahler potential.

Supersymmetric vacua are characterized by the vanishing of the F-term conditions,
\begin{eqnarray}
\label{DW}
D_{t_a} W = D_{N_k} W = D_{T_\lambda} W = 0 \,.
\end{eqnarray}
In this subsection we consider the general structure of these equations, and show that in general all geometric moduli can be fixed by fluxes.  We shall focus on the regime of large volumes, where a geometric description is possible; however as described above, these equations  can also be applied to the small volume region if the corrections are known.

\subsubsection{Complex structure equations}

The complex structure equations $D_{N_k} W = D_{T_\lambda} W = 0$ become
\begin{eqnarray}
\label{DWComplex}
p_k + 2 i \,e^{2D} \, W \, \Im (C g_k) = 0 \,, \\
q_\lambda + 2 i \,e^{2D}\, W \,\Im (C Z_\lambda) = 0 \,.
\end{eqnarray}
The first observation is that the imaginary part of each of these equations is identical.  Given that $C$ and $D$ are real, one simply finds
\begin{eqnarray}
\label{ComplexAxionEqn}
\Re W = q_\lambda \tilde\xi_\lambda - p_k \xi_k + \Re W^K = 0\,.
\end{eqnarray}
This turns out to be the unique condition from (\ref{DW}) involving the axions.  As a result, only a single linear combination of the $\xi_k$, $\tilde\xi_\lambda$ fields is fixed; the
remaining $\xi_k$, $\tilde\xi_\lambda$ fields are the only moduli that cannot be stabilized using fluxes.

This collapse of what was apparently $h^{2,1}+1$ constraints into a
single constraint can be traced to the fact that the constant
coefficients $p_k, q_\lambda$ are
real, and therefore do not contain enough degrees of freedom to
stabilize both the complex structure moduli and the associated axions.
A similar thing happens in the case of $G_2$ flux vacua \cite{ADV};
we compare these ensembles in section \ref{sec:discussion}.

We found the same result in section \ref{sec:moduli}, where the only constraint on the
axions arises from the space-time polarized $F_4$ through the
Chern-Simons term $\int H_3 \wedge C_3 \wedge F_4$.
This is not an issue in our $T^6/\bZ_3^2$ example because there $h^{2,1}=0$, so
the single constraint suffices to fix the single axion arising
from the dilaton multiplet.  In more general examples, Euclidean D2 instantons are expected to lift the
remaining axions \cite{amirtom,BBS}.  In fact in this general class of models, the allowed
$H_3$ fluxes live in the cohomology group $H^3_{-}$ while
the axions come from $H^3_{+}$. Hence the instantons which lift these axions are precisely the ones
allowed by both the orientifold projection and by the nontrivial fluxes.

Turning to the real parts of (\ref{DWComplex}), we note that $\Im W
=0$ is incompatible with any nonzero $H_3$ flux; we will see upon
studying the K\"ahler sector that $\Im W \neq 0$ when any RR fluxes
are turned on, as long as $\int J\wedge J \wedge J \neq 0$.  Given
nonzero $\Im W$, we find that if any $p_k$ or $q_\lambda$ vanishes,
the corresponding modulus $\Im g_k$ or $\Im Z_k$ must vanish.  Then
for any $k_i$ or $\lambda_j$ with nonzero $p_{k_i}$, $q_{\lambda_j}$, we can eliminate $e^D \, \Im W$
to obtain
\begin{eqnarray}
e^{-K^{\rm cs}/2}\, {p_{k_1} \over \Im g_{k_1}} = e^{-K^{\rm cs}/2}\, {p_{k_2} \over \Im g_{k_2}} = \ldots =  e^{-K^{\rm cs}/2} \, {q_{\lambda_1}
  \over \Im Z_{\lambda_1}} = \ldots  \equiv Q_0\,.
\end{eqnarray}
These equations are invariant under an overall rescaling of $\Omega$ and hence depend only on the inhomogeneous coordinates on the complex structure moduli space;
combined with the vanishing of $\Im g_k$ or $\Im Z_k$ for the cases when $p_k, q_\lambda = 0$, they constitute $h^{2,1}$ real equations that will in general fix all the $h^{2,1}$ complex structure moduli, independent of the RR fluxes or values of the K\"ahler moduli.  The final equation from (\ref{DWComplex}) can then be cast as
\begin{eqnarray}
\label{DilatonEqn}
e^{-\phi} = 4 \sqrt{2} \, e^{K^K/2} \,{ \Im W  \over Q_0} \,,
\end{eqnarray}
which determines the dilaton once the complex and K\"ahler moduli have
been solved for.

Before turning to the K\"ahler moduli, we derive a useful consequence of the complex structure equations.  Multiplying the equations (\ref{DWComplex}) by $\Re (C Z_k)$ and $\Re (C g_\lambda)$, respectively, summing over $k$ and $\lambda$ and taking the difference, we find using the identity (\ref{ComplexIdentity}) that
\begin{eqnarray}
-i W = \sum_\lambda q_\lambda \Re (C g_\lambda) - \sum_k p_k \Re (C Z_k) \equiv {1 \over 2} \Im W^Q \,.
\end{eqnarray}
Hence when the complex structure moduli satisfy their equations of motion, the vacuum value of the superpotential can be written in terms of the K\"ahler moduli only:
\begin{eqnarray}
\label{DecoupleEqn}
W(t_a, N_k, T_\lambda) = - i \Im W^K(t_a) \,.
\end{eqnarray}

\subsubsection{K\"ahler equations}

The relation (\ref{DecoupleEqn}) allows us to decouple the K\"ahler sector.  Using (\ref{DecoupleEqn}), the equations $D_{t_a} W =0$ become
\begin{eqnarray}
\label{DWKahler}
\partial_{t_a} W^K - i \partial_{t_a} K^K \Im W^K = 0 \,.
\end{eqnarray}
Hence we can consider these equations entirely independently from the hypermultiplet moduli and $H_3$ fluxes.

In the analysis that follows we will assume nonvanishing $m_0$.  It is straightforward to show that for $m_0 = 0$, one must either have $m_a = e_a = 0$ as well, and the K\"ahler moduli are then all unfixed, or the $v_a$ are driven to zero, far from the large-volume region.

Again it is useful first to consider the imaginary parts of the equations.  Since $K^K$ depends only on $v_a \equiv \Im t_a$, the second term in (\ref{DWKahler}) is real.  Thus we find
\begin{eqnarray}
\Im \partial_{t_a} W^K = \kappa_{abc} v_b ( m_c - m_0\, b_c) = 0 \,,
\end{eqnarray}
(recall $b_c  = \Re t_c$).  The regularity of the moduli space metric
implies there is always some $\kappa_{abc}$ that is nonzero for any
given $c$; assuming the 2-cycle volumes $v_b$ do not vanish, as will
be the case for example in a geometrical limit, one finds for all $c$:
\begin{eqnarray}
\label{KahlerAxion}
b_c = {m_c \over m_0} \,.
\end{eqnarray}
We see that unlike the case of the complex structure, for the K\"ahler
moduli the axions are generically all fixed.  As we will discuss
further
in section \ref{sec:discussion}, this can be understood as arising from
the fact that the K\"ahler sector has twice as many fluxes per real
modulus as the complex structure
($m^a, e_a$ for the K\"ahler sector as opposed to $p_q, q_\lambda$ for
the complex structure sector).

Consider now the real part of the equations (\ref{DWKahler}).  Using the axion solution (\ref{KahlerAxion}), one can write these equations as
\begin{eqnarray}
\label{KahlerEqns}
\left( 3 m_0^2 \kappa_{abc} v_b v_c + 4 e_a m_0 + 2 \kappa_{abc} m_b
m_c \right) (\kappa_{def} v_d v_e v_f) \hspace*{1in}\\
+(\kappa_{abc} v_b v_c) (6 m_0 e_d v_d + 3 \kappa_{def} m_d m_e v_f) = 0 \,.
\nonumber
\end{eqnarray}
Multiplying by $v_a$ and summing over $a$, we have
\begin{eqnarray}
\label{KahlerVolume}
3 m_0^2 (\kappa_{abc} v_a v_b v_c) +10 m_0 e_a v_a + 5 \kappa_{abc} m_a m_b v_c =0 \,.
\end{eqnarray}
Substituting this back into (\ref{KahlerEqns}) and cancelling an overall factor, one finds for each $a$,
\begin{eqnarray}
\label{KahlerSoln}
3 m_0^2 \kappa_{abc} v_b v_c + 10 m_0 e_a + 5 \kappa_{abc} m_b m_c =0\,.
\end{eqnarray}
These $h^{1,1}_-$ simple quadratic equations for the $h^{1,1}_-$ moduli $v_a$ are the final result; we have as many equations as unknowns and expect all the moduli to be frozen.  Let us discuss a few properties of these equations.

A key feature of (\ref{KahlerSoln}) is that K\"ahler moduli are only
coupled to other K\"ahler moduli with which they have a nonvanishing
triple intersection; this is not obvious from the original equations
(\ref{DWKahler}).  In studying our example $T^6/\bZ_3^2$ in section
\ref{sec:4D-applied}, we shall see that this justifies treating every
blow-up mode independently from the other blow-ups, as well as from
the untwisted moduli, even when the latter are not taken to be much larger than the blow-ups.

Using (\ref{KahlerVolume}), one can show that
\begin{eqnarray}
\label{VacuumW}
W = -i \Im W^K = {2 i \over 15} m_0 \kappa_{abc} v_a v_b v_c \,.
\end{eqnarray}
From this we learn that $W=0$ cannot occur for this class of vacua without the overall volume $\int J \wedge J \wedge J$ vanishing.
This justified the assumption of $\Im W \neq 0$ we made in analyzing the
complex structure moduli\footnote{A compactification with $m_0 \neq 0$ and
all other fluxes vanishing (and no orientifold) was studied in \cite{Polchinski-Strominger},
where it was found that the solution is forced to $\int J \wedge J \wedge J = 0$,
consistent with (\ref{KahlerVolume}).}.

Using (\ref{VacuumW}) one can solve for the dilaton using (\ref{DilatonEqn}).  One can see from (\ref{KahlerSoln}), (\ref{VacuumW}) that under a flip of the sign of all RR fluxes, $W \to - W$.  Thus to preserve the physically correct sign for the dilaton (\ref{DilatonEqn}), one must flip the signs of the $H_3$ fluxes as well. ( The periods $\Im g_k$ and $\Im Z_\lambda$ have definite sign fixed by the sign of $\Omega$, which in turn is fixed as it calibrates the special Lagrangian submanifolds on which the O6s are wrapped.)  It is familiar from studying type IIB vacua that flipping signs of the RR fluxes without doing likewise for the NSNS fluxes leads to a solution with unphysical dilaton, an indication that the solution preserves the opposite sign of supersymmetry; the sign of the tadpole from the fluxes has been flipped, and in this case, those fluxes are consistent with an anti-O6 background instead of an O6 background.

Let us summarize the equations determining the supersymmetric vacua.
\vspace*{0.1in}

\noindent K\"ahler  moduli $b_a, v_a$:

\begin{eqnarray}
\label{KahlerSummarySoln}
b_a = {m_a \over m_0} \,,\;\;\;\;\;
 \;\;\;\;\; \;\;\;\;\; 3 m_0^2 \kappa_{abc} v_b v_c + 10
m_0 e_a + 5 \kappa_{abc} m_b m_c =0\,.
\end{eqnarray}
Complex structure moduli $\Im g_k, \Re Z_\lambda$:
\begin{eqnarray}
\Im g_k = 0 \;  {\rm for} \; p_k = 0\,,
\;\;\;\;\; \;\;\;\;\; \Im Z_\lambda = 0 \; {\rm
  for} \; q_\lambda = 0
\,,
\end{eqnarray}
\begin{eqnarray}
e^{-K^{\rm cs}/2}\, {p_{k_1} \over \Im g_{k_1}} = e^{-K^{\rm cs}/2}\, {p_{k_2} \over \Im g_{k_2}} = \ldots =  e^{-K^{\rm cs}/2} \, {q_{\lambda_1}
  \over \Im Z_{\lambda_1}} = \ldots  \equiv Q_0\,, \;\; {\rm for \; all} \; p_{k_i,} q_{\lambda_j} \neq
  0\,.
  \nonumber
\end{eqnarray}
\noindent Dilaton $\phi$:
\begin{eqnarray}
e^{-\phi} = {4 \sqrt{2} \over 5 \sqrt{3}}\, {m_0 \over Q_0}  (\kappa_{abc} v^a v^b v^c)^{1/2} \,.
\end{eqnarray}
\noindent One axion $q_\lambda \tilde{\xi}_\lambda -p_k \xi_k$:
\begin{eqnarray}
p_k \xi_k - q_\lambda \tilde\xi_\lambda = \Re W^K = e_0 + {e_a m_a
  \over m_0} + { \kappa_{abc} m_a m_b m_c \over 3 m_0^2} \,.
\end{eqnarray}
These equations assume $v_a \neq 0$ and $\kappa_{abc} v_a v_b v_c \neq 0$.

Note that in general we need $m_0$ and at least one $p_k$ or
$q_\lambda$ to be nonzero for a stabilized vacuum; if either
condition fails, all fluxes must vanish and the moduli go
unstabilized.  The minimum set of fluxes required to stabilize all
geometric
moduli is $m_0$, one $p_k$ or $q_\lambda$ (satisfying the orientifold
tadpole) and one $e_a$ or $m_a$ for each K\"ahler modulus.

It will generally be true that some fluxes will lead to solutions of (\ref{KahlerSummarySoln}) lying outside the geometric regime; for example in section \ref{sec:4D-applied} we will see that for the $T^6/\bZ_3^2$ orientifold some fluxes imply some $v^a < 0$.  In this regime we expect not just the K\"ahler potential $K$, but also the superpotential $W$, to receive $\alpha'$ corrections, and hence the result cannot be trusted.

When some of the $p_k$ or $q_\lambda$ vanish, one ends up with either
$g_k =0$ or $Z_\lambda = 0$.
The vanishing of a linear combination of periods does not a priori
mean that a 3-cycle has collapsed; such vanishing occurs at a dense set
of points in moduli space, while the actual discriminant locus is of
codimension one.\footnote{We thank F. Denef for reminding us of this fact.}
In the rare case where such
a 3-cycle has collapsed, one might
worry about being driven to a singularity on moduli space where new
fields become light.  However in type IIA string theory, the complex
structure moduli space is embedded within the quaternionic
hypermultiplet moduli space, within which singularities have codimension
four or higher.  Even after the orientifold projection, since the surviving
axion partners of the complex moduli are in general unfixed by the
fluxes, one need
not end up at a singular point; landing at the singular point in
moduli space will require a tuning of the axion vevs.

\subsection{Gauge redundancies}
\label{sec:gauge}

There are in general modular group transformations, acting both on the moduli and on the fluxes, that relate equivalent vacua.  In particular, it is evident that there are two kinds of modular transformations of infinite order, those that shift the complex structure axions $\xi_k$, $\tilde\xi_\lambda$ by one, and those that shift the K\"ahler axions $b_a$ likewise.  Here we derive the action of these transformations on the fluxes; we do so using the fact that the $DW=0$ equations must transform covariantly, so solutions are mapped to other solutions.  We also heuristically describe the nature of the transformation in a T-dual type IIB picture.

Consider first the K\"ahler axions.  From (\ref{KahlerAxion}), is is obvious that a shift of the axion $t_a \to t_a +1$ corresponds to a shift of $m_a \to m_a + m_0$.  Assuming $m_0$ is fixed, the first term of (\ref{KahlerVolume}) is unchanged, thus determining the action on $e_a$.  Finally the invariance of $\Re W^K$ can be used to fix the transformation of $e_0$.  The result is, for integers $u_a$,
\begin{eqnarray}
\label{KahlerAxionShift}
t_a &\to& t_a + u_a \,, \\ m_0 \to m_0 \,, \quad m_a \to m_a + u_a m_0 \,, &&
 e_a \to e_a - \kappa_{abc} m_b u_c \,, \quad e_0 \to e_0 - e_a u_a \,.
\nonumber
\end{eqnarray}
This transformation can be regarded as the T-dual of a geometric transformation.  Consider the $(T^6)/\bZ_3^2$ model and a shift of $t \to t +1$ for one of the tori, which is a shift in $B_2$ integrated over that $T^2$.  Taking a single T-duality in this $T^2$, the shift of $t$ is mapped to trivial shift of the complex structure of the dual torus, while the RR fields are mapped into modes of $F_1$, $F_3$ and $F_5$, which are mixed amongst each other by this geometrical shift in precisely the way specifying (\ref{KahlerAxionShift}).

Next consider shifts of the complex structure axions.  Consider for example $\xi_k  \to \xi_k + 1$, which requires $\Re W^K \to \Re W^K + p_k$; this can be accomplished with a shift of $e_0$ alone.  In general we find
\begin{eqnarray}
\label{ComplexAxionShift}
\xi_k \to \xi_k + U_k \,, \quad \tilde\xi_\lambda \to \tilde\xi_\lambda + V_\lambda \,, \quad e_0 \to e_0 + p_k U_k - q_\lambda V_\lambda \,,
\end{eqnarray}
for integers $U_k$, $V_\lambda$.  When only one component of $H_3$ is turned on, this transformation can be understood as the mirror of type IIB $SL(2,Z)$ shifts; three T-dualities take $H_3$ and $F_6$ to type IIB $H_3$ and $F_3$ polarized along the same directions, which are then mixed by an $SL(2,Z)$ transformation.

\section{Application to $T^6/\bZ_3^2$ model}
\label{sec:4D-applied}

We now apply the results of the previous section to the specific case
of our $T^6/\bZ^3$ model, searching for solutions in the
limit where all volumes are sufficiently large that we can neglect
$\alpha'$ corrections.

We shall denote the $F_2$ and $F_4$ fluxes associated to the untwisted
cycles by $m_i$ and $e_i$, $i = 1,2,3$ while those on the blow-ups are
$n_A$ and $f_A$, $A = 1 \ldots 9$; the corresponding moduli are the
untwisted K\"ahler modes $t_i$ and the blow-up K\"ahler modes
$t_{B_A}$.
In the hypermultiplet sector, $h^{2,1} = 0$ and we have
only the index $k=0$ and no $\lambda$ indices; the unique flux is
$p_{k=0} \equiv p$, and the moduli are just the dilaton $\phi$ and its
axionic partner $\xi_{k=0} \equiv \xi$.

\subsection{General solution}

We first consider the K\"ahler sector.  The
nonzero elements of the intersection
form are $\kappa_{123} = \kappa$ and $\kappa_{AAA} = \beta$, and
consequently we can solve for each of
the blow-up modes
independently of the untwisted moduli and of the other blow-up
modes.\footnote{The values of $\kappa$ and $\beta$ can be found by a
simple modification of the results in \cite{Strominger}, where the
intersection form of $T^6/\bZ_3$ was computed.  The result (correcting
a minor error in \cite{Strominger} and accounting for the further
free $\bZ_3$ action) is that $\kappa = 81$ and $\beta = 9$, but we will
continue the discussion in terms of variable $\kappa, \beta$.}
Considering first the untwisted moduli, the axions are fixed as (\ref{KahlerAxion}),
\begin{eqnarray}
b_i = \Re t_i  = {m_i \over m_0} \,,
\end{eqnarray}
while for the volumes $v_i = \Im t_i$ we find the equations (\ref{KahlerSoln})
\begin{eqnarray}
6 m_0^2 \kappa  v_2 v_3 + 10 m_0 e_1 + 10 \kappa m_2 m_3 &=& 0 \,, \\
6 m_0^2 \kappa  v_1 v_3 + 10 m_0 e_2 + 10 \kappa m_1 m_3 &=& 0 \,, \\
6 m_0^2 \kappa  v_1 v_2 + 10 m_0 e_3 + 10 \kappa m_1 m_2 &=& 0 \,.
\end{eqnarray}
The solution to this system is
\begin{eqnarray}
\label{OurKahlerSoln}
v_i = {1 \over |\hat{e}_i|} \sqrt{{-5 \hat{e}_1 \hat{e}_2 \hat{e}_3 \over 3 m_0 \kappa} } \,,
\end{eqnarray}
where we have defined the shifted flux $\hat{e}_i$ invariant under shifts of $t_i$ (\ref{KahlerAxionShift}),
\begin{eqnarray}
\hat{e}_i \equiv e_i + {\kappa m_j m_k \over m_0} \,,
\end{eqnarray}
where $j$ and $k$ are simply the two values other than $i$.

For each of the blow-up modes, the volumes $v_{B_A}$ satisfy
\begin{eqnarray}
3 m_0^2 \beta v_{B_A}^2  + 10 m_0 f_A + 5 \beta n_A^2 = 0 \,,
\end{eqnarray}
with no sum over $A$.  The solution for the complex blow-up moduli is then\footnote{Note that to stay within the K\"ahler cone, one should
choose the solution with $\Im t_{B_A} < 0$; this unusual convention
arises because the self-intersection of the resolving $\bP^2$ of a
$\bC^3/\bZ_3$ singularity is $-3$ times an actual curve.}
\begin{eqnarray}
\label{OurBlowUpSoln}
t_{B_A} = {n_A \over m_0} - i \sqrt{{- 10 \hat{f}_A \over 3 \beta m_0}} \,.
\end{eqnarray}
where again we defined an invariant shifted flux $\hat{f}_A$,
\begin{eqnarray}
\hat{f}_A \equiv f_A + {\beta n_A^2 \over 2 m_0} \,.
\end{eqnarray}
There are no complex structure moduli, so only the dilaton and its axion $\xi$ remain.
Using the dilaton equation (\ref{DilatonEqn}), and the results $\Im g_0 = -1/\sqrt{2}$, $K^{\rm cs} = 0$, $e^{K^K} = 3/(4 \kappa_{abc} v_a v_b v_c)$, we find
\begin{eqnarray}
\label{OurDilSoln}
e^{-\phi} = - {4 \sqrt{3} \over 15} {m_0 \over p} (\kappa_{abc} v_a v_b v_c)^{1/2} \,,
\end{eqnarray}
where the total volume (proportional to the 4D coupling $e^{-D}$) is given by
\begin{eqnarray}
\kappa_{abc} v_a v_b v_c =  - {15 p \over 2 \sqrt{2} m_0} e^{-D} = {10 \over |m_0|} \sqrt{-5 \hat{e}_1 \hat{e}_2 \hat{e}_3 \over 3 m_0 \kappa} + \beta \sum_A \left( {-10 \hat{f}_A \over 3 \beta m_0 }\right)^{3/2}\,,
\end{eqnarray}
where we have used the fact, discussed in the next subsection, that $\sgn (m_0 \hat{e}_1 \hat{e}_2 \hat{e}_3) < 0$ must hold.
Finally the axion $\xi$ is fixed as (\ref{ComplexAxionEqn})
\begin{eqnarray}
\xi = {\Re W^K \over p} = {1 \over p} \left(  e_0 + {e_i m_i + f_A n_A \over m_0} + { 6 \kappa m_1 m_2 m_3 + \beta \sum_A n_A^3 \over 3 m_0^2}\right) \,.
\end{eqnarray}

\subsection{Regime of validity and agreement with 10D analysis}
\label{sec:validity}

This solution will be valid as long as the volumes $v_i$, $v_{B_A}$ are sufficiently large that $\alpha'$ corrections can be neglected, and the string coupling is small enough that quantum corrections can be neglected.  One can see from (\ref{OurKahlerSoln}) and (\ref{OurBlowUpSoln}) that the volumes are large whenever
\begin{eqnarray}
|\hat{e}_i| \gg |m_0| \,, \quad \quad |\hat{f}_A| \gg |m_0| \,.
\end{eqnarray}
Moreover, to remain within the K\"ahler cone, we must ensure the untwisted volumes are sufficiently larger than the blow-ups, requiring
\begin{eqnarray}
\label{FluxHierarchy}
|\hat{e}_i| \gg |\hat{f}_A| \gg |m_0| \,.
\end{eqnarray}
Because the four-form and two-form fluxes are not constrained by the tadpole, we are free to scale them to be as large as we wish.  Thus we can always choose some fluxes obeying (\ref{FluxHierarchy}) that provide a geometric solution.

When the hierarchy (\ref{FluxHierarchy}) is obeyed, the behavior of physical quantities is dominated by the $F_4$ flux for the non-blow up cycles.  Let us again take $\hat{e}_i \sim \bar{e}$; we have shown that the K\"ahler parameters scale as $v_i \sim \bar{e}^{1/2}$, becoming large with large $\bar{e}$.  Then in addition to the overall volume becoming big, the ten- and four-dimensional string couplings become small in this limit:
\begin{eqnarray}
\label{Scalings}
\vol \sim \bar{e}^{3/2} \,, \quad \quad e^{\phi} \sim \bar{e}^{-3/4} \,, \quad \quad e^{D} \sim \bar{e}^{-3/2} \,,
\end{eqnarray}
suppressing quantum corrections.

One may be concerned that even though the volumes are much larger than $\alpha'$, higher derivative corrections to the 10D Lagrangian may nonetheless become relevant, because the flux parameter $\bar{e}$ will increase the coefficient of certain terms as it grows large.  We can estimate the size of higher order corrections involving powers of $|F_4|^2$ as follows.

First, two powers of $F_4$ give an explicit $\bar{e}^2$ scaling.  Next there are 4 factors of the inverse metric in contracting the indices of the form fields, which provides a factor of $R^{-8} \sim \bar{e}^{-2}$. Finally, it is a famous fact that RR vertex operators are accompanied by an extra factor of $g_s$, yielding an additional power of $\bar{e}^{-3/2}$.

Assembling all of the ingredients, we see that
relative to the leading term in the 10D Lagrangian, terms with additional
powers of $|F_4|^2$ are suppressed by an expansion parameter
$\lambda \sim \bar{e}^{-3/2}$.  Therefore, in the large $\bar{e}$ limit, we expect corrections from both the $\alpha^\prime$ and $g_s$ expansions to be parametrically
suppressed.  The existence of these SUSY vacua is therefore
robust against any known corrections.

The scalings (\ref{Scalings}) are the same as those found in the 10D
analysis; in fact, in the
limit (\ref{FluxHierarchy}), where the fluxes on the non-blow-up
cycles dominate the string coupling, the solution
 (\ref{OurKahlerSoln}), (\ref{OurDilSoln}) agrees precisely with
(\ref{eq:TenDSoln}), and the blow-up volume (\ref{OurBlowUpSoln}) agrees
qualitatively with the estimate (\ref{TenDBlowUp}), with the replacement $\hat{e}_i \to e_i$,
$\hat{f}_A \to f_A$ to reflect the special case $m_i = n_A = 0$.
There is one subtlety: the signs in the 4D analysis are more
constrained than those in the 10D analysis.  In particular, although
both analyses agree that a solution requires
\begin{eqnarray}
\label{SignConstraints}
{\rm sgn}\, (m_0 p) < 0  \,,
\end{eqnarray}
the 4D supersymmetric equations also imply a constraint on the signs of the $F_4$ fluxes,
\begin{eqnarray}
{\rm sgn}\, (m_0 \hat{e}_1 \hat{e}_2 \hat{e}_3) < 0\,, \quad \quad  \sgn (m_0 \hat{f}_A ) < 0 \,,
\end{eqnarray}
as well as the condition
\begin{eqnarray}
\hat{e}_i v_1 = \hat{e}_2 v_2 = \hat{e}_3 v_3 \,,
\end{eqnarray}
requiring the signs of the $\hat{e}_i$ all to coincide:
\begin{eqnarray}
\label{SignE}
\sgn \hat{e}_1 = \sgn \hat{e}_2 = \sgn \hat{e}_3 \,,
\end{eqnarray}
in order for the $v_i$ to all be positive and hence in the large-volume region.  The results of section \ref{sec:10D}, however, imply that even if the signs of the $\hat{e}_i$ are not aligned,
there is still a solution at positive $v_i$, necessarily non-supersymmetric as it violates (\ref{SignE}), but apparently lacking in instabilities.  The nature of these extra solutions, and the exact location of the supersymmetric vacuum in the small volume region, we leave for future work.

Since $W \neq 0$, these supersymmetric vacua are anti-de Sitter.  Hence another interesting quantity to consider is the 4D cosmological constant $\Lambda$.  One finds
\begin{eqnarray}
\Lambda = - 3 e^{K^K + K^Q} |W|^2 \sim \bar{e}^{-9/2} \,.
\end{eqnarray}
It is natural to ask whether the vacuum can be treated as effectively four-dimensional: this will be the case if the Hubble scale $H$, defined as
\begin{eqnarray}
H^2 = {\Lambda \over M_P^2} \,,
\end{eqnarray}
with $M_P^2$ the four-dimensional Planck scale, is less than the Kaluza-Klein scale $1/R$.
Using the four-dimensional Einstein frame where $M_P \sim \bar{e}^0$, we calculate that $R^2 \sim \bar{e}^{7/2}$, leading to the result
\begin{eqnarray}
(H R)^2 \sim {1 \over \bar{e}} \,.
\end{eqnarray}
Hence there is a parametric hierarchy between the AdS radius and the Kaluza-Klein scale, and treating the vacuum with four-dimensional effective theory makes sense.  This is to be contrasted with the case of the Freund-Rubin vacua which feature most prominently in examples of the AdS/CFT correspondence, where the KK scale and the scale of the cosmological constant are the same, and the background is not effectively four-dimensional.

Hence we have demonstrated for the $T^6/\bZ_3^2$ orientifold the existence of parametrically tunable  large volume, weak coupling flux vacua with a valid four-dimensional description and all moduli stabilized.

\section{Rudimentary IIA vacuum statistics}
\label{sec:discussion}

\begin{quote}
``To understand God's thoughts we must study statistics, for these are
the measure of His purpose."

\hspace{3in} --- {\it Florence Nightingale}
\end{quote}

It would be interesting to do a thorough analysis of the statistics of
type IIA flux vacua; related M-theory models were recently studied in
\cite{ADV}.  Here, we make a modest contribution by analyzing the
statistics in the simplest toy model, a fictitious rigid Calabi-Yau
space with a single K\"ahler modulus $t$ and no complex structure
moduli, with fluxes $m_0$, $m$, $e$, $e_0$ and $p$.  Taking $\kappa = 1$ we have
\begin{equation}
W^K = e_0 + e\,t +\frac{1}{2} m\,t^2 -\frac{m_0}{6} \, t^3 \,.
\end{equation}
This example may
be viewed as somewhat analogous to the type IIB rigid Calabi-Yau toy model
studied in \cite{statrefs}.

The solution for the K\"ahler modulus $t$ is identical to that of (\ref{OurBlowUpSoln}) for a single blow-up mode,\footnote{Unlike in (\ref{OurBlowUpSoln}), here we have chosen the positive root, assuming $v$ is the volume of a physical curve.}
\begin{eqnarray}
\label{ToySoln}
t = {m \over m_0} + i \sqrt{-10 \hat{e}\over3 m_0} \,, \quad \quad \hat{e} \equiv e + {m^2 \over 2 m_0} \,,
\end{eqnarray}
while the dilaton and axion have the solutions
\begin{eqnarray}
\label{ToySoln2}
e^{-\phi} \sim {m_0 \over p} (\Im t)^3 \,, \quad \xi = {1 \over p} ( e_0 + {e m \over m_0} + {m^3 \over 3 m_0^2}) \,.
\end{eqnarray}
While we have chosen to analyze this case for simplicity, it is easy to see that the solutions for $T^6/\bZ_3^2$ are virtually identical when $e_1 = e_2 = e_3$, $m_1= m_2 = m_3$.  In fact, because of the simple form of the K\"ahler equations (\ref{KahlerSoln}), all solutions for K\"ahler moduli in the geometric regime will have the general form (\ref{ToySoln}).  Hence we are able to capture the essential behavior of all flux-frozen K\"ahler moduli by studying (\ref{ToySoln}).

The tadpole condition in general requires
\begin{equation}
\label{o6can}
0 < - m_0 p \leq N \,,
\end{equation}
where $N$ is the magnitude of the negative D6 charge induced by
O6 planes wrapping the fixed point locus of the
anti-holomorphic involution $\sigma$.  In the cases that the
inequality is not saturated, one can compensate by including D6 branes.
In simple examples (including our explicit case) $N$ is ${\cal O}(1)$,
and we shall assume some fixed (though possibly large) $N$
in the following analysis ({\it i.e.}
we will ${\it not}$ work in an $N \to \infty$ limit).
 In a more general model with some complex structure moduli,
there will be a tadpole like (\ref{o6can}) for each $p$ or $q$,
 limiting the possible values of all the NSNS fluxes.

By varying flux integers, it appears that one can easily obtain
a denumerably infinite number of vacua. However, the naive analysis significantly over-counts solutions, as there are modular symmetries that relate vacua with different values of these parameters.  While we will only do approximate statistics
of various asymptotics in an appropriate large flux limit, to avoid
making a serious error we need to gauge fix the modular group generators
of infinite order.  These are the two symmetries discussed in section (\ref{sec:gauge}): integer shifts of $\Re t$ and of $\xi$, which we account for as follows.

Rather than restricting the value of the moduli with the gauge symmetries, we choose instead to restrict possible choices of the fluxes.  Using the shift symmetry (\ref{KahlerAxionShift}) of the K\"ahler axions, we can restrict
\begin{eqnarray}
0 \leq  m < \vert m_0 \vert \,,
\end{eqnarray}
leaving us with $\vert m_0 \vert$
inequivalent choices of the flux $m$ for fixed $m_0$; in more complicated models there will be one such symmetry for each $m_a$, permitting them all to be restricted in this way.
We can estimate the number of such choices as
\begin{equation}
\label{divisorsum}
\sum_{M=1}^{N} \sum_{m_0 | M} \vert m_0 \vert = \sum_{M=1}^{N}
\sigma(M)
\sim \frac{\pi^2}{ 12} N^2 ~.
\end{equation}
We can then fix the shift symmetry (\ref{ComplexAxionShift}) of the axion $\xi$ by restricting $e_0$:
\begin{eqnarray}
\label{EZeroFix}
0 \leq e_0 < p \,,
\end{eqnarray}
giving us $p$ possible values.  In fact, this is not independent of the previous discussion: for a given partition of $M \leq N$ into $m_0 p$, one gets $m_0$ choices
for $m$ and $p$ choices for $e_0$, so we should replace
(\ref{divisorsum}) by the slightly more elaborate
\begin{equation}
\label{truecount}
\sum_{M=1}^{N} \sum_{m_0 | M} M = \sum_{M=1}^N M d(M)
\sim N^2 \log N \,.
\end{equation}
Notice that in models with multiple $\xi$ axions, further gauge fixing beyond the restriction (\ref{EZeroFix}) will be necessary.

At this point we can see that for a given orientifold, $m_0$ and $p$ are constrained to take a finite number of values, and the degeneracy of vacua from varying $m$ and $e_0$ is given by (\ref{truecount}); almost all the fluxes have been restricted to finitely many values.   However, we are still free to vary $e$ while satisfying all tadpole conditions, and we have no more infinite order modular symmetries to
reduce the space of choices to a finite set.
Furthermore, we see from (\ref{ToySoln}), (\ref{ToySoln2}) and
that if we are concerned with the large $e$ asymptotics of
the solutions (as we will be), then the allowed variations of
$m_0, m$ at fixed $N$ will have a minor effect.
In the explicit example, for instance, $N = {\cal O}(1)$ and
the additional degeneracy factors discussed here are completely
irrelevant for understanding the distribution of vacua at large
$e$.

The upshot of this discussion is that, in this gauge fixing,
the statistics are dominated by the large $e$ vacua and
we shall focus henceforth on their properties.

\subsection{Statistics and general comments}
\label{sec:statistics}

Although the number of vacua diverges, there are still interesting
statistical questions that one can ask.
The well-posed questions are questions like: how many vacua exist
below a given volume?  How many vacua exist above a given $\vert \Lambda
\vert$?

It is easy to answer these questions using the scaling results of
the previous subsection; essentially all that matters is the large
$e$ behavior, since this is where an infinite number of vacua lie,
with their properties dominated by $e$.
The finite range of values of the other fluxes then only contributes
to very fine structure in the space of vacua.

Using the fact that the length scale of the compact space in string frame scales as $R \sim e^{1/4}$, we see that (at least for sufficiently large $R^*$) the number of
vacua with $R \leq R^*$ scales like $(R^*)^4$:
\begin{equation}
\label{raddist}
{\cal N} (R \leq R^*) \sim (R^*)^4~.
\end{equation}
In previous cases, Calabi-Yau flux vacua have had distributions
governed by the volume form on the appropriate moduli space; we note
here that (\ref{raddist}) does ${\it not}$ conform to a distribution
on the K\"ahler moduli space governed by the volume form arising from
(\ref{eq:kahlerpot}).

For the cosmological constant, using $\Lambda \sim e^{-9/2}$, one has
\begin{equation}
\label{ccdist}
{\cal N} (|\Lambda| \geq |\Lambda^*|) \sim (|\Lambda^*|)^{-2/9}~.
\end{equation}
In other words, the number distributions of vacua (without any
assumptions about a cosmological measure) favor large volume and
small cosmological constant, in this supersymmetric ensemble.
Note that one should not trust the distribution (\ref{ccdist}) at
large $|\Lambda|$ because our approximations are invalid at
small $e$.  Hence the slow power of the decay in this limit
should not cause concern;
any structure in the distribution of vacua at ${\it large}$
$|\Lambda|$ is not trustworthy.

Given the large amount of recent work on characterizing the string
landscape, it seems worthwhile to make some comments about the
similarities and differences of our results to those obtained in
other ensembles.
Firstly, we should emphasize that the divergence of the number of
SUSY vacua may not be particularly disastrous.  A mild cut on the
acceptable volume of the extra dimensions will render the number
of vacua finite.  On the other hand, one can legitimately worry that
the conclusions of any statistical argument will be dominated
by the precise choice of the cut-off criterion, since the
regulated distribution is dominated by vacua with volumes close
to the cut-off.

Secondly, we should comment that our statistical results are qualitatively
rather similar to those obtained in \cite{ADV} for $AdS_4$ Freund-Rubin
vacua of M-theory.
A promising
difference between these two sets of vacua is the
parametric ratio we obtain
between the Hubble scale and the scale of the internal
dimensions, which is generally absent in Freund-Rubin vacua.

\subsection{Comparison to other ensembles of vacua}

By far the most well-studied example of flux vacua in string theory is the set of type IIB vacua with the Calabi-Yau complex structure moduli and dilaton stabilized by $H_3$ and $F_3$ fluxes.  In addition, recently there has been discussion \cite{ADV} of statistics for moduli-stabilized flux vacua in compactifications of M-theory on manifolds of $G_2$ holonomy.  It is naturally interesting to compare the ensembles to the IIA system we study.

In principle any vacuum of string theory can be described in an
alternate duality frame, and so the vacua we describe should be
expressible in the language of type IIB string theory via mirror
symmetry, or of M-theory by relating the string coupling to the M-theory circle.
However, our vacua need not admit a description
as a flux compactification in the dual language.  In fact, generically
some parameters associated to fluxes will be mapped to geometric
torsions, which are considerably more difficult to characterize; an
understanding of them on the same level as fluxes has yet to be
obtained \cite{Generalized, duals}.  Furthermore, the global properties of the dual-spaces may even be nongeometric \cite{Nongeo}.  Hence by studying flux compactifications of a
given theory, without torsions, we are choosing a different ``slice"
of all possible compactifications than we would obtain by studying the
flux compactifications of another theory.

So while by studying torsions as well as fluxes we could in principle see that two dual descriptions of string theory have the same vacuum statistics,\footnote{This is not guaranteed, however, especially if one only computes the
statistics for those vacua which are weakly coupled in the respective
corners (which may be the sensible thing to do).} different ensembles of flux compactifications alone will not in general agree.  Hence it is interesting to compare them.

We will take a small step in this direction by comparing the ratio of available fluxes to moduli in four different ensembles: Type IIA K\"ahler, type IIA complex structure, type IIB complex structure, and M-theory on $G_2$.  Define the ratio
\begin{eqnarray}
\eta \equiv {\# \;{\rm real \; flux \; parameters} \over \# \;{\rm real \; moduli}} \,.
\end{eqnarray}
In general the larger the $\eta$ parameter is for a given ensemble,
the more moduli can be fixed, and the less
``friendly" the distribution will be (in the language of \cite{ADK});
similar observations have been put forward in \cite{Douglas,ADV}.

The simplest ensemble is the case of M-theory on a $G_2$ manifold \cite{Bobby, ADV}.
There are $b_3$ complex moduli $z_i$, and $b_3$ $G_4$ fluxes $N^i$, as well as the complex
 Chern-Simons invariant $c_1 + i c_2$.
 Hence we have $\eta_{G_2} = (b_3 + 2)/(2b_3) \sim 1/2$.
  The superpotential has the structure
\begin{eqnarray}
\label{WGtwo}
W_{G_2} = c_1 + i c_2  + z_i N^i \,.
\end{eqnarray}
In this ensemble, nonzero $c_2$ is required for solutions at finite volumes $s_i = \Im z_i$, and only a single linear combination of the axions $\Re z_i$ are fixed.  We may understand this heuristically as since $\eta \sim 1/2$, there are only as many fluxes as there are volume parameters $s_i$, and consequently the axions are left unfixed.

Consider next the type IIA K\"ahler sector studied in this paper; since it can be completely decoupled from the other moduli, it makes sense to consider it independently.  There are $h^{1,1}_-$ complex moduli $t_a$, and $2 h^{1,1}_- + 2$ RR fluxes; hence we find $\eta_{IIA,K} = (2 h^{1,1}_- + 2) / (2 h^{1,1}_-) \sim 1$.  The superpotential
\begin{eqnarray}
\label{WIIAK}
W_{IIA,K} = e_0 + t_a e_a + {1 \over 2} \kappa_{abc} m_a t_b t_c - {m_0 \over 6} \kappa_{abc} t_a t_b t_c \,,
\end{eqnarray}
is structurally a generalization of (\ref{WGtwo}), with the fluxes $m_a$ and $m_0$ generating quadratic and cubic terms.  With this doubling of the number of fluxes, one finds that the axions as well as the geometric moduli are stabilized.

Hence one sees how in passing from an M-theory description
to a IIA description, additional parameters that were described
in terms of the geometry have become available as fluxes,
and the increase in the number of fluxes allows all moduli
to be stabilized.  Note that (\ref{WIIAK}) has no precise
analog of $c_2$ in (\ref{WGtwo}),
the Chern-Simons invariant introduced by Acharya \cite{Bobby}
to achieve nontrivial moduli stabilization, but $m_0$ plays
a very similar role.

Next consider the other ensemble in type IIA compactifications, that of the complex structure moduli and dilaton.  There are $h^{2,1}+1$ complex moduli, and in addition to the $h^{2,1} + 1$ $H_3$ fluxes, one requires the complex number $\Im W^K$ from the K\"ahler sector as input.  Hence we have $\eta_{IIA,c} = (h^{2,1} + 3) / (2(h^{2,1} + 1)) \sim 1/2$.  Since one has $\eta_{IIA,c} = \eta_{G_2}$, one might expect a similar story, and this is what we found: as in the $G_2$ case, the geometric moduli are frozen, but the axions are not except for a single linear combination.  Hence we see that although the $G_2$ superpotential superficially resembles the IIA K\"ahler case more strongly (they are both simple polynomials in the moduli), its behavior is much more like that of the IIA complex structure case,
and this similarity can be traced to their having the same value of $\eta$.

The final familiar ensemble is that of type IIB, with imaginary
self-dual fluxes stabilizing the complex structure moduli and dilaton.
In this case there $h^{2,1}+1$ complex moduli, but $4(h^{2,1}+1)$
fluxes; hence $\eta_{IIB} = 4 (h^{2,1} +1) / (2(h^{2,1} +1)) = 2$.
This is the largest number of fluxes per modulus of all these ensembles; one may
think of starting with the type IIA complex structure ensemble and
doubling the fluxes, as $F_3$ contributes as well as $H_3$,
effectively complexifying the flux.  (Of course, since in IIB the RR
fluxes as well as the NSNS fluxes go into stabilizing the complex
structure moduli, there are none left to stabilize the K\"ahler
moduli.)  Not only are all moduli frozen, but additional choices are
left over, allowing a broader, less ``friendly" distribution.

Thinking ahead, the inclusion of torsions as well as fluxes will naturally cause the suitable generalization of the $\eta$ parameter to grow.  Hence when one considers all the discrete choices that characterize these generalized flux compactifications, stabilization of all moduli becomes increasingly easy, and distributions become less and less ``friendly".  It is quite reasonable to expect that a generic example of such a generalized flux compactification would stabilize all moduli, regardless of the particular string theory considered.

\section{Conclusions}

The main striking features of
the class of models described in this paper are their simplicity, and
the appearance of a parameter which yields power-law parametric control.
In the supersymmetric vacua of the IIB theory where all moduli
are stabilized \cite{KKLT}, the control parameter only grows
logarithmically with a tuning parameter; hence, while one can make
controlled constructions, it requires precise tuning in a large
space of flux vacua.  Here,
in contrast, the radii and couplings fall into a controlled regime
as a power of the $F_4$ flux.  This gives these models special appeal
as a setting to do controlled studies of fully stabilized string vacua.
It also hints that finding dual CFTs, which is a difficult problem
for the AdS models of \cite{KKLT}, may be considerably simpler here;
the large flux limit may admit a simple dual description.

It would be worthwhile to find proposals for perturbing
these vacua by small positive energies to yield controlled de Sitter
models, perhaps along the lines of similar proposals in the type IIB
theory \cite{KKLT, dSLift}.  In addition, the inclusion of perturbative corrections to
$K$, worldsheet instantons
(whose effects should be computable by using mirror symmetry and co-opting the
appropriate type IIB computations of prepotentials), and
Euclidean D2 instanton effects, could add very interesting
features to these potentials; in the analogous ${\cal N}=2$ setting
quantum corrections certainly do seem to play an important role \cite{KK}.
At least the worldsheet instanton effects should be something
that one can incorporate at the level of statistical analyses.
There has also been great progress in constructing realistic
brane world models in flux backgrounds \cite{review} and in using
the fluxes to freeze the open string moduli \cite{uranga,open}
and induce soft supersymmetry breaking terms \cite{softsusy};
it would be interesting
to combine these ingredients in the setting suggested here.

Finally, it would be interesting to see if there is a direct relation between
our IIA constructions and some topological field theory construction,
which could provide an analogue of the Hartle-Hawking wavefunction \cite{HH}
for these vacua  -- such a construction has been obtained for
some simple classes of Freund-Rubin vacua in \cite{OVV}.
We note here that any naive application of the Hartle-Hawking wavefunction
to obtain a measure on this set of vacua will suffer from the same problem
of cut-off dominance mentioned in \S6.1\ in the context of
statistical arguments.  Without imposing a cut-off
on the four-form flux, the wavefunction will be badly non-normalizable
(as it is for the analogous black hole problem in \cite{OVV}, if one
does not impose a cut-off on the allowed charges).  Imposing a cut-off,
one will find that the wavefunction is peaked at the cut-off; this is
the analogue of the cut-off domination problem for statistical arguments.
One proposal to fix this problem in the more physical case of de Sitter vacua
has been described in the papers \cite{Tye}, which also provide references
to further critical discussion in the quantum cosmology literature.
At any rate it is clearly a worthwhile and ambitious goal to find a
good measure on the space of vacua.  Success will require
both a detailed knowledge of the structure of the space of vacua, and
significant new insights
into early universe cosmology in string theory.

\begin{center}
\bf{Acknowledgements}
\end{center}
\medskip
We are grateful to B.~Acharya, P.~Aspinwall, T.~Banks, D.~Belov, F.~Denef,
M.~Dine, M.~Douglas, D.~Freedman, T.~Grimm, A.~Hanany,
A.~Kashani-Poor, B.~K\"ors, H.~Liu, L.~McAllister, J.~McGreevy,
A.~Micu, R.~Myers, Y.~Okawa, R.~Roiban, J.~Shelton, S. ~Shenker and
E.~Silverstein for discussions and useful comments. We are
particularly grateful to F.~Denef for a thorough reading of the
manuscript.
Thanks also to G.\ Moore for helpful comments and discussion regarding
topological aspects of flux quantization and Chern-Simons terms.
O.~D.\ would like to thank the University of Toronto and
the Perimeter Institute where some of this work was done.  The work of
O.~D.\ was supported by NSF grant PHY-0243680. The work of A.~G.\ was
supported in part by INTAS 03-51-6346. The work of S.~K.\ was
supported by a David and Lucile Packard Foundation Fellowship for
Science and Engineering, the D.O.E. under contract DE-AC02-76SF00515,
and the National Science Foundation under grant 0244728. The work of
W.~T.\ was supported by the DOE under contract \#DE-FC02-94ER40818.

\vspace*{0.2in}

\appendix

\section{IIA Chern-Simons term in presence of fluxes}
\label{sec:appendix}
\vspace*{0.1in}

In this appendix we consider the Chern-Simons term of IIA supergravity
in the presence of topological fluxes.  We only consider the massless
IIA theory here, which can be derived through dimensional reduction
from M-theory, and give an elementary derivation of that subset of the full set of Chern-Simons terms that plays a role in the analysis of this paper.
A full treatment of the Chern-Simons terms of type IIA string theory is rather subtle and requires dealing properly with flux self-duality, anomaly cancellation and the classification of fluxes in K-theory \cite{CS}, and leads to additional contributions involving curvature forms and an overall sign for the exponentiated action; we neglect such terms here.

The Chern-Simons term of IIA supergravity is well known to be given in
the absence of topological fluxes by
\begin{equation}
S_{\rm CS} = \frac{1}{2 \kappa_{10}^2} \int  H_3 \wedge C_3 \wedge F_4 \,.
\label{eq:CS1}
\end{equation}
In the absence of topological fluxes, this Chern-Simons term can be
integrated by parts to give
\begin{equation}
S_{\rm CS} = -\frac{1}{2 \kappa_{10}^2} \int B_2 \wedge F_4 \wedge F_4 \,.
\label{eq:CS2}
\end{equation}
These two forms of the Chern-Simons contribution to the action are
generally used interchangeably.  Note, however, that in the presence
of a topological flux $H_3^{\rm bg}$ or $F_4^{\rm bg}$ there is a
subtlety.  When such a flux is present, the boundary terms
$\int_\partial B_2 \wedge C_3 \wedge F_4$ arising from the integration
by parts may not vanish, due to a large gauge transformation which
relates the forms $B_2, C_3$ at two images of the same boundary.
Thus, the two Chern-Simons terms (\ref{eq:CS1}, \ref{eq:CS2}) are not
necessarily equivalent in the presence of topological background
fluxes. In fact, if we decompose  $B^{{\rm total}} = B^{{\rm bg}} +
 B$ and $C_{3}^{\rm total}= C_{3}^{\rm bg} +  C_{3}$, we see
that the problem arises from taking either $B^{\rm bg}$ or $C_3^{\rm
  bg}$ to appear without a derivative in the action.  In this case,
the action is not necessarily gauge invariant under large gauge
transformations.

As an explicit example of this problem, consider by analogy a simple
U(1) gauge theory on a cubic $T^3$ with sides of length 1, with
connection $A_i$ and field strength $F_{ij} =
\partial_iA_j-\partial_jA_i$.  In this model the Chern-Simons term
$\int A \wedge F$ is invariant under local gauge transformations.  The
topological flux $F_{ij}$ is quantized to be $F_{ij} = 2 \pi n_{ij},
n_{ij} \in\bZ$.  Let us turn on an explicit $F_{12}$ flux by setting
$A_2^{\rm bg} = 2 \pi x_1$, and compute the term in the action which
gives a tadpole in this background to the fluctuation $A_3 = \lambda
\cos 2 \pi x_1\rightarrow F_{13} = -2 \pi \lambda\sin 2 \pi x_1$.  This
tadpole arises from the term
\begin{eqnarray}
\int A_2 F_{31} & = & \int_0^1dx_1 \; (2 \pi)^2 x_1 \lambda \sin 2 \pi
 x_1\\
 & = &  -2 \pi \lambda \,.
\end{eqnarray}
We might try integrating this term by parts, in which case we get a
boundary contribution
\begin{eqnarray}
\int A_2^{\rm bg} (\partial_3 A_1 -\partial_1 A_3) & \rightarrow &
-\int A_2^{\rm bg} \partial_1 A_3
\\
 & = &  \int F_{12}^{\rm bg} A_3 - \left( A_2^{\rm bg} A_3 \right)
|^1_0\\
& = & -2 \pi A_3 (x_1 = 1) = -2 \pi \lambda \,.
\end{eqnarray}
Thus, the integration by parts is not valid here if the boundary term
is neglected.  Furthermore, if we perform the global gauge
transformation
\begin{equation}
A_i \rightarrow A_i -i g^{-1} \partial_ig \,,
\end{equation}
where
\begin{equation}
g = e^{-2 \pi ix_1 x_2} \,,
\end{equation}
 we have
\begin{eqnarray}
A_1^{\rm bg} & = &  -2 \pi x_2 \,,\\
A_2^{\rm bg} & = &  0 \,.
\end{eqnarray}
The tadpole for the fluctuation $F_{13} = 2 \pi \lambda \sin 2 \pi x_1$
in this background explicitly vanishes!  Thus, the action $\int A
\wedge F$ is not invariant under large gauge transformations when the
background topological flux is encoded in $A$ which appears explicitly
without derivatives in the action.

To avoid these complications, we need to find an invariant definition
of the Chern-Simons term in the presence of topological fluxes.
A
correct definition of a $D$-dimensional Chern-Simons term $\Gamma$ on
a manifold $M_D$ is given by finding a $(D + 1)$-dimensional manifold
$ M_{D +1}$ with boundary $M_D = \partial M_{D +1}$.  Then
\begin{equation}
\int_{M_D} \Gamma = \int_{M_{D +1}} d \Gamma \,,
\end{equation}
is gauge invariant under {\it all} gauge transformations on $M_D$
which can be extended to gauge transformations on $M_{D +1}$ as long
as $d \Gamma$ is gauge invariant.  Note that generally $\Gamma$
depends on a $p$-form potential $C$, and not just on $dC$, so that
$\Gamma$ must be extended to $M_{D +1}$ by extending $C$ and not $dC$.
We will use this approach to find the invariant definition of the
Chern-Simons term in M-theory, which we then reduce to type IIA.  A
similar discussion of the Chern-Simons term of M-theory was given in
\cite{WittenM}.

To construct the Chern-Simons term of M-theory, we begin by making the
simplifying assumption that we have an $M_{11}$ which decomposes
as $M_{11} =\bR \times \hat{M}_{10}$ ,
such that there is no topological
flux of the M-theory 4-form $F_{\mu \nu \rho \sigma}$ with an index on
the first dimension.  We define $F^{\rm total} = dC + F^{\rm bg}$.
We can then write $M_{11} = \partial M_{12}$
where $M_{11} =H_+ \times \hat{M}_{10}$
with $H_+$ the upper half-plane.
We can then extend any $C_3$ from $M_{11}$ to $M_{12}$ by
multiplying by a function of the extra coordinate which is 1 on the
boundary and goes to 0 sufficiently rapidly in the interior.  For
example we could take
$ e^{-r}$ on
$H_+$.
We extend $F^{\rm bg}$ trivially on $M_{12}$, which amounts to
choosing a particular representative $F^{\rm bg} = dC^{\rm bg}$ and
extending $C^{\rm bg}$ trivially (though note that $C^{\rm bg}$ may
transform nontrivially between charts covering $M_{11}$).
The
four-form flux in 12D is then given by
\begin{equation}
\tilde{F} = d (e^{-r}C) + F^{\rm bg} = -dr \wedge e^{-r} C + e^{-r} F +
F^{\rm bg} \,.
\end{equation}
We can then directly integrate
\begin{eqnarray}
\int_{M_{12}}  \tilde{F}_4^{\rm total}\wedge  \tilde{F}_4^{\rm
  total}\wedge \tilde{F}_4^{\rm
  total}
   & \rightarrow &  \int_{0}^{\infty} dr \wedge (e^{-r}C) \wedge (e^{-r} F +
F^{\rm bg})\wedge (e^{-r} F +
F^{\rm bg})\nonumber\hspace*{0.2in}\\
 & = &  \frac13 C \wedge F \wedge F
+  C \wedge F \wedge F^{\rm bg} + C \wedge F^{\rm bg} \wedge F^{\rm bg} \,.
\end{eqnarray}
The coefficient of the first
term is fixed to agree with the term in the absence of
background fluxes, so that using conventions of Polchinski we have
\begin{equation}
S^{11}_{{\rm CS}} =
-\frac{1}{12 \kappa_{11}^2}
\int_{M_{11}} C_{3} \wedge
\left(F_4^{\rm } \wedge F_4^{\rm }
+3 F_4^{\rm } \wedge F_4^{\rm bg}
+3F_4^{\rm bg} \wedge F_4^{\rm bg}\right)  \,.
\end{equation}
This fixes the Chern-Simons term of M-theory in the presence of
background fluxes as long as there is a trivial one-dimensional factor
in $M_{11}$.

Now, we can dimensionally reduce to 10 dimensions.  Following the
standard dimensional reduction as in \cite{Polchinski} but using our conventions for RR fields, we have
\begin{eqnarray}
S^{{\rm IIA}}_{{\rm CS}} & = &
-\frac{1}{2 \kappa_{10}^2} \int \left[
B_2 \wedge F_4 \wedge F_4
+ 2B_2 \wedge F_4 \wedge F_4^{\rm bg} + C_3 \wedge H_3^{\rm bg} \wedge F_4
\right.\label{eq:CS-IIA}\\
 &  & \hspace*{1in} + \left.B_2 \wedge F_4^{\rm bg} \wedge F_4^{\rm bg}
+ 2 C_3 \wedge H_3^{\rm bg} \wedge F_4^{\rm bg} \right] \,, \nonumber
\end{eqnarray}
where we have integrated by parts where possible.
Note that these terms reduce correctly to (\ref{eq:CS1}, \ref{eq:CS2})
in the absence of topological fluxes.  The first line of
(\ref{eq:CS-IIA}) contains all terms needed in the case of
compactification of IIA on a 6-dimensional manifold, where there are
no terms quadratic in the topological background flux, since this
would require a nontrivial cohomology cycle of degree 7 or higher.  In
this case, which is the case of interest in this paper, the
Chern-Simons terms are precisely those found in \cite{KK} to be
compatible with the structure imposed by 4D supergravity.


\begin{thebibliography}{10}

\bibitem{KKLT}
S. Kachru, R. Kallosh, A. Linde and S. Trivedi, ``de Sitter vacua in
string theory,'' Phys. Rev. {\bf D68} (2003) 046005 [arXiv:hep-th/0301240].

\bibitem{KKLTexample}
F. Denef, M. Douglas, B. Florea, A. Grassi and S. Kachru, ``Fixing
all moduli in a simple F-theory compactification,'' [arXiv:hep-th/0503124];
F. Denef, M. Douglas and B. Florea, ``Building a better racetrack,''
JHEP {\bf 0406} (2004) 034 [arXiv:hep-th/0404257];
P. Aspinwall and R. Kallosh, to appear.

\bibitem{nonsusyKKLT}
J. Conlon, F. Quevedo and K. Suruliz, ``Large-volume flux compactifications:
Moduli spectrum and D3/D7 soft supersymmetry breaking,''
[arXiv:hep-th/0505076];
P. Berglund and P. Mayr, ``Non-perturbative superpotentials in F-theory
and string duality,'' [arXiv:hep-th/0504058];
V. Balasubramanian, P. Berglund, J. Conlon and F. Quevedo, ``Systematics
of moduli stabilisation in Calabi-Yau flux compactifications,''
JHEP {\bf 0503} (2005) 007 [arXiv:hep-th/0502058].


\bibitem{Eva}
A. Saltman and E. Silverstein, ``A new handle on de Sitter
compactifications,'' [arXiv:hep-th/0411271];
A. Maloney, E. Silverstein and A. Strominger, ``de Sitter
space in noncritical string theory,'' [arXiv:hep-th/0205316].

\bibitem{Bobby}
B. Acharya, ``A moduli fixing mechanism in M-theory,''
[arXiv:hep-th/0212294].

\bibitem{Carlos}
B. de Carlos, A. Lukas and S. Morris, ``Non-perturbative vacua
for M-theory on G2 manifolds,'' JHEP {\bf 0412} (2004) 018
[arXiv:hep-th/0409255.]


\bibitem{Heterotic}
G. Curio, A. Krause and D. L\"ust, ``Moduli stabilization in the
heterotic/IIB discretuum,'' [arXiv:hep-th/0502168];
S. Gurrieri, A. Lukas and A. Micu, ``Heterotic on Half-flat,''
Phys.Rev. {\bf D70} (2004) 126009 [arXiv:hep-th/0408121];
K. Becker, M. Becker, K. Dasgupta, P. Green and E. Sharpe, ``Compactifications
of Heterotic Strings on Non-Kahler complex manifolds II,''
Nucl.Phys. {\bf B678} (2004) 19 [arXiv:hep-th/0310058];
M. Becker, G. Curio and A. Krause, ``De Sitter Vacua from
Heterotic M Theory,'' Nucl.Phys. {\bf B693} (2004) 223 [arXiv:hep-th/0403027];
R. Brustein and S.P. de Alwis, ``Moduli Potentials in String
Compactifications with Fluxes: Mapping the Discretuum,''
Phys.Rev. {\bf D69} (2004) 126006 [arXiv:hep-th/0402088];
S. Gukov, S. Kachru, X. Liu and L. McAllister,
``Heterotic Moduli Stabilization with Fractional Chern-Simons Invariants,''
Phys.Rev. {\bf D69} (2004) 086008 [arXiv:hep-th/0310159];
E. Buchbinder and B. Ovrut, ``Vacuum Stability in Heterotic M-theory,''
Phys.Rev. {\bf D69} (2004) 086010 [arXiv:hep-th/0310112];
G. Cardoso, G. Curio, G. Dall'Agata and D. L\"ust, ``Heterotic string
theory on non-Kahler manifolds with H-flux and gaugino condensate,''
Fortsch. Phys. {\bf 52} (2004) 483 [arXiv:hep-th/0310021];
G. Cardoso, G. Curio, G. Dall'Agata and D. L\"ust, ``BPS action
and superpotential for heterotic string compactifications with
fluxes,'' JHEP {\bf 0310} (2003) 004 [arXiv:hep-th/0306088].

\bibitem{Antoniadis}
  I.~Antoniadis and T.~Maillard,
``Moduli stabilization from magnetic fluxes in type I string theory,''
  Nucl.\ Phys.\ B {\bf 716}, 3 (2005)
  [arXiv:hep-th/0412008];
  I.~Antoniadis, A.~Kumar and T.~Maillard,
``Moduli stabilization with open and closed string fluxes,''
[arXiv:hep-th/0505260].



\bibitem{BP}
R. Bousso and J. Polchinski, ``Quantization of four-form fluxes and
dynamical neutralization of the cosmological constant,'' JHEP {\bf
006} (2000) 006 [arXiv:hep-th/0004134];
J. Brown and C. Teitelboim,
``Dynamical neutralization of the cosmological constant,''
Phys.Lett. {\bf B195} (1987) 177;
J. Brown and C. Teitelboim,
``Neutralization of the cosmological constant by membrane creation,''
Nucl. Phys. {\bf B297} (1988) 787;
J. Feng, J. March-Russell, S. Sethi and F. Wilczek,
``Neutralization of the cosmological constant by membrane creation,''
Nucl. Phys. {\bf B602} (2001) 307 [arXiv:hep-th/0005276];
L. Susskind,
``The anthropic landscape of string theory,''
[arXiv:hep-th/0302219].

\bibitem{Douglas}
M. Douglas, ``The statistics of string/M theory vacua,''
JHEP {\bf 0305} (2003) 046 [arXiv:hep-th/0303194].

\bibitem{statrefs}
S. Ashok and M. Douglas, ``Counting flux vacua,'' JHEP {\bf 0401} (2004)
060 [arXiv:hep-th/0307049];
F. Denef and M. Douglas, ``Distributions of flux vacua,'' JHEP {\bf 0405}
(2004) 072 [arXiv:hep-th/0404116];
O. DeWolfe, A. Giryavets, S. Kachru and W. Taylor, ``Enumerating flux
vacua with enhanced symmetries,'' JHEP {\bf 0502}
(2005) 037 [arXiv:hep-th/0411061].

\bibitem{statrefsII}
A. Giryavets, S. Kachru and P. Tripathy, ``On the taxonomy of flux vacua,''
JHEP {\bf 0408} (2004) 002 [arXiv:hep-th/0404243];
M.~Dine, E.~Gorbatov and S.~Thomas,
  ``Low energy supersymmetry from the landscape,''
  arXiv:hep-th/0407043;
J. Conlon and F. Quevedo, ``On the explicit construction and statistics
of Calabi-Yau flux vacua,'' JHEP {\bf 0410} (2004) 039 [arXiv:hep-th/0409215];
J. Kumar and J. Wells, ``Landscape cartography: A coarse survey of
gauge group rank and stabilization of the proton,'' Phys. Rev. {\bf D71}
(2005) 026009 [arXiv:hep-th0409218];
M.~Dine,
  ``Supersymmetry, naturalness and the landscape,''
  arXiv:hep-th/0410201;
R. Blumenhagen, F. Gmeiner, G. Honecker, D. L\"ust and T. Weigand,
``The statistics of supersymmetric D-brane models,'' Nucl. Phys. {\bf B713}
(2005) 83 [arXiv:hep-th/0411173];
M.~Dine, D.~O'Neil and Z.~Sun,
  ``Branches of the landscape,''
  arXiv:hep-th/0501214.




\bibitem{Grimm-Louis-IIA}
T.~Grimm and J.~Louis,
``The effective action of type IIA Calabi-Yau orientifolds,''
[arXiv:hep-th/0412277].


\bibitem{KK}
S.~Kachru and A.~Kashani-Poor,
``Moduli potentials in type IIA compactifications with RR and NS flux,''
JHEP {\bf 0503} (2005) 066 [arXiv:hep-th/0411279].


\bibitem{Zwirner}
J.~Derendinger, C.~Kounnas, P.~Petropoulos and F.~Zwirner,
``Superpotentials in IIA compactifications with general fluxes,''
[arXiv:hep-th/0411276];
G.~Villadoro and F.~Zwirner,
``N = 1 effective potential from dual type-IIA D6/O6 orientifolds with general
arXiv:hep-th/0503169;
J.~P.~Derendinger, C.~Kounnas, P.~M.~Petropoulos and F.~Zwirner,
``Fluxes and gaugings: N = 1 effective superpotentials,''
arXiv:hep-th/0503229.

\bibitem{others}
G.~Dall'Agata and N.~Prezas,
``N = 1 geometries for M-theory and type IIA strings with fluxes,''
Phys.\ Rev.\ D {\bf 69}, 066004 (2004)
[arXiv:hep-th/0311146];
D. L\"ust and D. Tsimpis, ``Supersymmetric AdS(4) compactifications
of IIA supergravity,'' JHEP {\bf 0502} (2005) 027 [arXiv:hep-th/0412250];
K. Behrndt and M. Cvetic, ``General ${\cal N}=1$ supersymmetric fluxes
in massive type IIA string theory,'' Nucl. Phys. {\bf B708} (2005) 45
[arXiv:hep-th/0407263];
K. Behrndt and M. Cvetic, ``General ${\cal N}=1$ supersymmetric flux
vacua of (massive) type IIA string theory,'' [arXiv:hep-th/0403049];
J. Gauntlett, D. Martelli and D. Waldram, ``Superstrings with intrinsic
torsion,'' Phys. Rev. {\bf D69} (2004) 086002 [arXiv:hep-th/0302158].

\bibitem{Generalized}
M. Grana, R. Minasian, M. Petrini and A. Tomasiello, ``Supersymmetric
backgrounds from generalized Calabi-Yau manifolds,'' JHEP {\bf 0408} (2004)
046 [arXiv:hep-th/0406137];
M. Grana, R. Minasian, M. Petrini and A. Tomasiello, ``Type II strings
and generalized Calabi-Yau manifolds,'' Comptes Rendus Physique {\bf 5}
(2004) 979 [arXiv:hep-th/0409176].




\bibitem{orbifold}
L. Dixon, J. Harvey, C. Vafa and E. Witten, ``Strings on Orbifolds,''
Nucl. Phys. {\bf B261} (1985) 678.

\bibitem{Strominger}
A. Strominger,
``Topology of superstring compactification,''
in {\it Unified String Theories}, M.\ Green and D.\ Gross eds., World
Scientific (1986).

\bibitem{bgk}
R. Blumenhagen, L. Gorlich and B. Kors,
``Supersymmetric 4D orientifolds of type IIA with D6-branes at angles,''
JHEP {\bf 0001} (2000) 040 [arXiv:hep-th/9912204].



\bibitem{Romans}
L.J. Romans, ``Massive ${\cal N}=2A$ supergravity in ten-dimensions,''
Phys. Lett. {\bf B169} (1986) 374.

\bibitem{Polchinski}
J. Polchinski,
{\it String theory. Vol. 2: Superstring theory and beyond},
Cambridge University Press, 1998.

\bibitem{KTheory}
R.~Minasian and G.~W.~Moore,
  ``K-theory and Ramond-Ramond charge,''
  JHEP {\bf 9711}, 002 (1997)
  [arXiv:hep-th/9710230];
E.~Witten,
 ``D-branes and K-theory,''
  JHEP {\bf 9812}, 019 (1998)
  [arXiv:hep-th/9810188];
G.~W.~Moore and E.~Witten,
  ``Self-duality, Ramond-Ramond fields, and K-theory,''
  JHEP {\bf 0005}, 032 (2000)
  [arXiv:hep-th/9912279];
  D.~S.~Freed and M.~J.~Hopkins,
  ``On Ramond-Ramond fields and K-theory,''
  JHEP {\bf 0005}, 044 (2000)
  [arXiv:hep-th/0002027].


\bibitem{CS}
  D.~E.~Diaconescu, G.~W.~Moore and E.~Witten,
  ``E(8) gauge theory, and a derivation of K-theory from M-theory,''
  Adv.\ Theor.\ Math.\ Phys.\  {\bf 6}, 1031 (2003)
  [arXiv:hep-th/0005090];
  D.~E.~Diaconescu, G.~W.~Moore and E.~Witten,
  ``A derivation of K-theory from M-theory,''
  arXiv:hep-th/0005091;
  G.~W.~Moore and N.~Saulina,
  ``T-duality, and the K-theoretic partition function of typeIIA  superstring
  theory,''
  Nucl.\ Phys.\ B {\bf 670}, 27 (2003)
  [arXiv:hep-th/0206092];
  E.~Diaconescu, G.~W.~Moore and D.~S.~Freed,
  ``The M-theory 3-form and E(8) gauge theory,''
  arXiv:hep-th/0312069.


\bibitem{WittenM}
E.~Witten,
  ``On flux quantization in M-theory and the effective action,''
  J.\ Geom.\ Phys.\  {\bf 22}, 1 (1997)
  [arXiv:hep-th/9609122].

\bibitem{Beasley-Witten}
C. Beasley and E. Witten,
``A note on fluxes and superpotentials in M-theory
compactifications on  manifolds of G(2) holonomy,''
JHEP {\bf 0207} (2002) 046 [arXiv:hep-th/0203061].

\bibitem{BBS}
K.~Becker, M.~Becker and A.~Strominger,
 ``Five-branes, membranes and nonperturbative string theory,''
   Nucl.\ Phys.\ B {\bf 456}, 130 (1995)
     [arXiv:hep-th/9507158].

\bibitem{BF}
P.~Breitenlohner and D.~Z.~Freedman,
 ``Positive Energy In Anti-De Sitter Backgrounds And Gauged Extended Supergravity,''
  Phys.\ Lett.\ B {\bf 115}, 197 (1982);
P.~Breitenlohner and D.~Z.~Freedman,
``Stability In Gauged Extended Supergravity,''
  Annals Phys.\  {\bf 144}, 249 (1982).

\bibitem{Ganor-Sonnenschein}
O. Ganor and J. Sonnenschein,
``On the strong coupling dynamics of heterotic string theory on  C**3/Z(3),''
JHEP {\bf 0205} (2002) 018 [arXiv:hep-th/0202206].



\bibitem{Louis}
J. Louis and A. Micu,
``Type II theories compactified on Calabi-Yau threefolds in
the presence  of background fluxes,''
Nucl.Phys. {\bf B635} (2002) 395 [arXiv:hep-th/0202168];
T. Grimm and J. Louis,
``The effective action of N = 1 Calabi-Yau orientifolds,''
Nucl.Phys. {\bf B699} (2004) 387 [arXiv:hep-th/0403067].

\bibitem{Fluxpot}
S.~Gukov, C.~Vafa and E.~Witten,
``CFT's from Calabi-Yau four-folds,''
Nucl.\ Phys.\ B {\bf 584}, 69 (2000)
[Erratum-ibid.\ B {\bf 608}, 477 (2001)]
[arXiv:hep-th/9906070].
S.~Gukov,
``Solitons, superpotentials and calibrations,''
Nucl.\ Phys.\ B {\bf 574}, 169 (2000)
[arXiv:hep-th/9911011];
T.~R.~Taylor and C.~Vafa,
``RR flux on Calabi-Yau and partial supersymmetry breaking,''
Phys.\ Lett.\ B {\bf 474}, 130 (2000)
[arXiv:hep-th/9912152].


\bibitem{ADV}
B. Acharya, F. Denef and R. Valandro,
``Statistics of M theory vacua,''
[arXiv:hep-th/0502060].


\bibitem{amirtom}
A. Kashani-Poor and A. Tomasiello, ``A stringy test of flux-induced
isometry gauging,'' [arXiv:hep-th/0505208].







\bibitem{ADK}
N. Arkani-Hamed, S. Dimopoulos and S. Kachru,
``Predictive landscapes and new physics at a TeV,''
[arXiv:hep-th/0501082].


\bibitem{Polchinski-Strominger}
J. Polchinski and A. Strominger,
``New Vacua for Type II String Theory,''
Phys.Lett. {\bf B388} (1996) 736 [arXiv:hep-th/9510227].




\bibitem{duals}
S. Gurrieri, J. Louis, A. Micu and D. Waldram, ``Mirror symmetry
in generalized Calabi-Yau compactifications,'' Nucl. Phys.
{\bf B654} (2003) 61 [arXiv: hep-th/0211102];
S. Kachru, M. Schulz, P. Tripathy and S. Trivedi, ``New supersymmetric
string compactifications,'' JHEP {\bf 0303} (2003) 061 [arXiv:hep-th/0211182];
M. Schulz, ``Calabi-Yau duals of torus orientifolds,'' [arXiv:hep-th/0412270].

\bibitem{Nongeo}
S.~Hellerman, J.~McGreevy and B.~Williams,
  ``Geometric constructions of nongeometric string theories,''
  JHEP {\bf 0401}, 024 (2004)
  [arXiv:hep-th/0208174];
  A.~Flournoy, B.~Wecht and B.~Williams,
  ``Constructing nongeometric vacua in string theory,''
  Nucl.\ Phys.\ B {\bf 706}, 127 (2005)
  [arXiv:hep-th/0404217].

 \bibitem{dSLift}
 C.~P.~Burgess, R.~Kallosh and F.~Quevedo,
  ``de Sitter string vacua from supersymmetric D-terms,''
  JHEP {\bf 0310}, 056 (2003)
  [arXiv:hep-th/0309187];
  A.~Saltman and E.~Silverstein,
  ``The scaling of the no-scale potential and de Sitter model building,''
  JHEP {\bf 0411}, 066 (2004)
  [arXiv:hep-th/0402135].

\bibitem{review}
R. Blumenhagen, M. Cvetic, P. Langacker and G. Shiu, ``Toward
realistic intersecting D-brane models,'' [arXiv:hep-th/0502005].

\bibitem{uranga}
J. Cascales and A. Uranga, ``Branes on generalized calibrated
submanifolds,'' JHEP {\bf 0411} (2004) 083 [arXiv:hep-th/0407132].

\bibitem{open}
R. Blumenhagen, M. Cvetic, F. Marchesano and G. Shiu, ``Chiral
D-brane models with frozen open string moduli,'' JHEP {\bf 0503} (2005)
050 [arXiv:hep-th/0502095].

\bibitem{softsusy}
K. Choi, A. Falkowski, H. Nilles and M. Olechowski, ``Soft supersymmetry
breaking in KKLT flux compactification,'' [arXiv:hep-th/0503216];
D. L\"ust, P. Mayr, S. Reffert and S. Stieberger, ``F-theory flux,
destabilization of orientifolds and soft terms on D7-branes,''
[arXiv:hep-th/0501139];
F. Marchesano, G. Shiu and L. Wang, ``Model building and
phenomenology of flux-induced supersymmetry breaking on
D3-branes,'' Nucl. Phys. {\bf B712} (2005) 20 [arXiv:hep-th/0411080];
D. L\"ust, S. Reffert and S. Stieberger, ``MSSM with soft susy
breaking terms from D7-branes,'' [arXiv:hep-th/0410074];
L. Ibanez, ``The fluxed MSSM,'' Phys. Rev. {\bf D71} (2005) 055005
[arXiv:hep-ph/0408064];
P. Camara, L. Ibanez and A. Uranga, ``Flux-induced susy-breaking
soft terms on D7-D3 brane systems,'' Nucl. Phys. {\bf B708} (2005) 268
[arXiv:hep-th/0408036];
D. L\"ust, S. Reffert and S. Stieberger, ``Flux-induced soft supersymmetry
breaking in chiral type IIB orientifolds with D3/D7 branes,''
Nucl. Phys. {\bf B706} (2005) 3 [arXiv:hep-th/0406092];
A. Lawrence and J. McGreevy, ``Local string models of soft
supersymmetry breaking,'' JHEP {\bf 0406} (2004) 007 [arXiv:hep-th/0401034];
M. Grana, T. Grimm, H. Jockers and J. Louis, ``Soft supersymmetry
breaking in Calabi-Yau orientifolds with D-branes and fluxes,''
Nucl. Phys. {\bf B690} (2004) 21 [arXiv:hep-th/0312232].

\bibitem{HH}
J. Hartle and S. Hawking,
``Wave function of the Universe,''
Phys. Rev. {\bf D28} (1983) 2960.

\bibitem{OVV}
H. Ooguri, C. Vafa and E. Verlinde,
``Hartle-Hawking wavefunction for flux compactifications,''
[arXiv:hep-th/0502211].

\bibitem{Tye}
S. Sarangi and S.H. Tye, ``The boundedness of euclidean gravity and
the wavefunction of the universe,'' [arXiv:hep-th/0505104];
H. Firouzjahi, S. Sarangi and S.H. Tye, ``Spontaneous creation of
inflationary universes and the cosmic landscape,'' JHEP {\bf 0409} (2004)
060 [arXiv:hep-th/0406107].







\end{thebibliography}
\end{document}